\documentclass{aa}

\usepackage{times}
\usepackage{graphicx}
\usepackage{natbib}
\usepackage{mathptm}
\usepackage{dcolumn}

\newcolumntype{.}{D{.}{.}{-1}}

\begin{document}

\title{The evolution of faint AGN between z\,$\simeq$\,1 and
       z\,$\simeq$\,5 from the COMBO-17 survey}

\author{C. Wolf\inst{1,2} \and L. Wisotzki\inst{3,4}
\and A. Borch\inst{2} \and S. Dye\inst{2,5} 
\and M. Kleinheinrich\inst{2,6} \and K. Meisenheimer\inst{2} }

\institute{ Department of Physics, Denys Wilkinson Bldg.,
            University of Oxford, Keble Road, Oxford, OX1 3RH, U.K. 
       \and Max-Planck-Institut f\"ur Astronomie, K\"onigstuhl 17,
            D-69117 Heidelberg, Germany 
       \and Astrophysikalisches Institut Potsdam, 
            An der Sternwarte 16, D-14482 Potsdam, Germany
       \and Universit\"at Potsdam, Institut f\"ur Physik,
            Am Neuen Palais 10, D-14469 Potsdam, Germany
       \and Astrophysics Group, Blackett Lab,
            Imperial College, Prince Consort Road, London, U.K.  
       \and IAEF, Universit\"at Bonn, 
            Auf dem H\"ugel 71, D-53121 Bonn, Germany }

\date{Received / Accepted }

\abstract{
We present a determination of the optical/UV AGN luminosity function and 
its evolution, based on a large sample of faint ($R < 24$) QSOs identified 
in the COMBO-17 survey. Using multi-band photometry in 17 filters
within $350\,\mathrm{nm} \la \lambda_\mathrm{obs} \la 930\,\mathrm{nm}$,
we could simultaneously determine photometric redshifts with an 
accuracy of $\sigma_z<0.03$ and obtain spectral energy distributions.
The redshift range covered by the sample is $1.2 < z < 4.8$,
which implies that even at $z \simeq 3$, the sample reaches below 
luminosities corresponding to $M_B = -23$, conventionally employed 
to distinguish between Seyfert galaxies and quasars.
We clearly detect a broad plateau-like maximum of quasar activity around $z
\simeq 2$ and map out the smooth turnover between $z\simeq 1$ and $z\simeq 4$. 
The shape of the LF is characterised by some mild curvature, but no
sharp `break' is present within the range of luminosities covered.
Using only the COMBO-17 data, the evolving LF can be adequately described 
by either a pure density evolution (PDE) or a pure luminosity evolution 
(PLE) model. However, the absence of a strong $L^*$-like feature in the shape 
of the LF inhibits a robust distinction between these modes. 
We present a robust estimate for the integrated UV luminosity generation 
by AGN as a function of redshift. We find that the LF continues to rise 
even at the lowest luminosities probed by our survey, but that the slope 
is sufficiently shallow that the contribution of low-luminosity AGN 
to the UV luminosity density is negligible.
Although our sample reaches much fainter flux levels than previous 
data sets, our results on space densities and LF slopes are completely
consistent with extrapolations from recent major surveys such as SDSS and 2QZ.
}
\titlerunning{Evolution of faint AGN from COMBO-17}
\authorrunning{Wolf et al.}
\maketitle

\section{Introduction}

The luminosity function of quasi-stellar objects (QSOs) and its evolution
with redshift provides one of the most important tools for the cosmic 
demography of active galactic nuclei (AGN). It constrains physical models 
for QSOs, particularly those for the growth of supermassive black holes in 
galaxies within the context of hierachical collapse of structure in the 
universe \citep{HR93,Hai98,KH00}. It is also relevant for
understanding the extragalactic UV background \citep{MM93,BT98}.

Previous studies of the QSO luminosity function (QLF) have established a 
strong rise in the activity with look back time from the local universe 
to redshifts $z\sim2$ \citep{Boy88}. The main debate at these low to
intermediate redshifts is now about the question whether the shape of the 
QLF changes with redshift \citep{Boy00,Wis00,Miy00,Cow03}. 
If so, this would mean that low-luminosity AGN evolve differently 
from high-luminosity QSOs, possibly calling for a substantial revision 
of our understanding of the cosmic duty cycle of nuclear activity in galaxies.

At higher redshift $z>2$, the obvious expectation that the rise 
of the QSO activity with redshift must turn over at some point was 
satisfied by observations at $z>3$ \citep[hereafter WHO and SSG]{WHO94,SSG95}.
However, optical surveys in this redshift regime were often plagued
by selection effects and small number statistics. Moreover,
studies of X-ray-selected \citep{Miy00} and radio-selected QSOs 
\citep{JR00} did not strongly support claims of a very steep drop in 
nuclear activity for $z>3$, keeping the issue unresolved whether 
the detected turnover was physical reality or not.
A rather robust observation of the space density decline 
beyond $z>3.6$ has been established recently,
albeit only for very luminous QSOs,
by the Sloan Digital Sky Survey \citep[SDSS, ][]{Fan01}. 

Most optically selected QSO samples are expected to be more or less complete 
at either $z\la 2.2$ or $z\ga 3.6$, where QSOs show conspicuous colours in 
broad-band searches, while within this redshift band, confusion arises with 
stars and compact low-redshift galaxies. Any practical approach of 
following up QSO samples in this intermediate redshift range is forced to 
avoid strong contamination and observed mostly extreme objects, resulting
in a high degree of incompleteness that is difficult to account for. 
Thus, while the fact that cosmic QSO activity shows a maximum around 
$z\simeq 2$--3 is not doubted as such, this maximum has very rarely
been detected in a single survey.

This selection issue was one of the original motivations for the COMBO-17
survey. This survey is based on photometric classification and redshift 
estimation using a set of 17 filters, of which 12 are medium-band filters 
with $\sim10,000$~km/sec FWHM, well matched to optimum detection of 
QSO emission lines and spread across the range of wavelengths observable
by modern CCD detectors.

Another major driver for this project was to probe fainter regimes of 
the AGN luminosity function, a task which has always beed limited by the
overwhelming need for spectroscopic follow-up telescope time.
Using medium-band spectrophotometry, a reasonably complete and clean AGN 
sample could be obtained across a wide range of redshifts down to $R\la 24$. 
Based on such a deep sample, we were able to derive luminosity functions 
down to $M_B \la -23$ at redshifts above 3, entering the domain 
of Seyfert galaxies even at high redshift.

In this paper, we present a new determination of the luminosity function
of faint optically selected QSOs, aimed at a broad redshift range around the
elusive turnover. The analysis is based on a sample of 192 objects between 
$z=1.2$ and $z=4.8$, all selected from the COMBO-17 survey.
The survey is briefly described in Sect.~2, followed by a discussion of
technical aspects such as completeness and redshift quality in Sect.~3. 
In Sections 4 through 7 we present and discuss the results.

\section{COMBO-17 observations}

\begin{figure*}
\centering
\includegraphics[clip,angle=270,width=15.5cm]{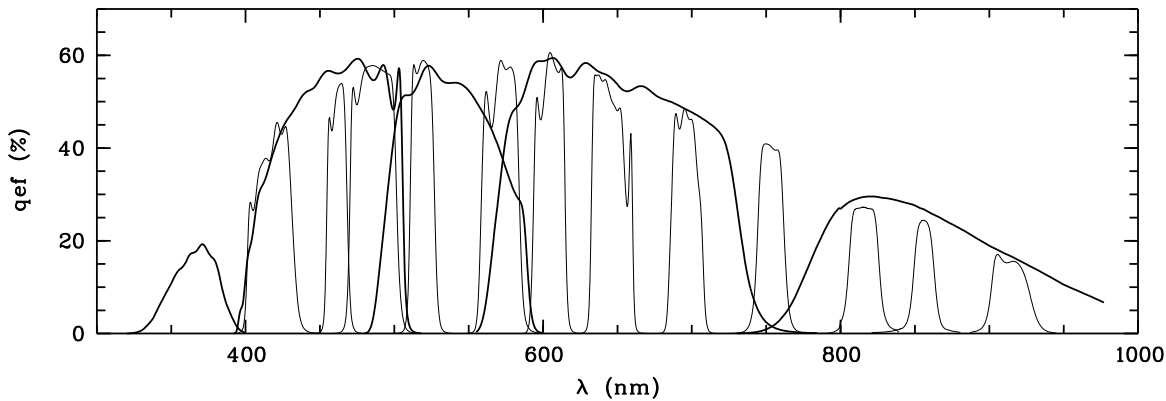}
\caption{COMBO-17 filter set: Total system efficiencies are shown in the 
COMBO-17 passbands, including two telescope mirrors, camera, CCD 
detector and average La Silla atmosphere. Combining all observations provides 
a low-resolution spectrum for all objects in the field.}
\label{qeff}
\end{figure*}

\begin{table}
\caption{The COMBO-17 filter set: Exposure times and 10$\sigma$ magnitude 
limits reached for point sources, averaged over all three fields. The $R$-band 
observations were selected to be taken under the best seeing conditions (FWHM 
$0\farcs55\ldots0\farcs8$). See also Fig.\ \ref{qeff}.
\label{filterset}}
\begin{tabular}{llrc}
\hline \hline
\multicolumn{2}{l}{$\lambda_\mathrm{\mathrm{cen}}$/fwhm (nm)}  & $t_\mathrm{\mathrm{exp}}$/sec & 
 $m_\mathrm{\mathrm{lim},10\sigma}$ \\ 
\noalign{\smallskip} \hline \noalign{\smallskip} 
364/38 &  $U$ &  20000 & 23.7 \\ 
456/99 &  $B$ &  14000 & 25.5 \\ 
540/89 &  $V$ &   6000 & 24.4  \\
652/162 & $R$ &  20000 & 25.2 \\ 
850/150 & $I$ &   7500 & 23.0 \\
\noalign{\smallskip} 
420/30  &      &  8000 & 24.0 \\ 
462/14  &      & 10000 & 24.0 \\ 
485/31  &      &  5000 & 23.8 \\ 
518/16  &      &  6000 & 23.6 \\ 
571/25  &      &  4000 & 23.4 \\ 
604/21  &      &  5000 & 23.4 \\ 
646/27  &      &  4500 & 22.7 \\
696/20  &      &  6000 & 22.8 \\ 
753/18  &      &  8000 & 22.5 \\ 
815/20  &      & 20000 & 22.8 \\ 
856/14  &      & 15000 & 21.8 \\ 
914/27  &      & 15000 & 22.0 \\
\noalign{\smallskip} \hline
\end{tabular}
\end{table}

\begin{table}
\caption{Positions and galactic reddening \citep{SFD98} for the 
three COMBO-17 fields analysed. All observations were obtained 
at the Wide Field Imager at the MPG/ESO 2.2\,m-telescope at La Silla. 
\label{fields} }
\begin{tabular}{llllll}
\hline \hline
Field   & $\alpha_\mathrm{\mathrm{J2000}}$ & $\delta_\mathrm{\mathrm{J2000}}$ & 
        $l_\mathrm{\mathrm{gal}}$ & $b_\mathrm{\mathrm{gal}}$ & $E_\mathrm{B-V}$ \\ 
\noalign{\smallskip} \hline \noalign{\smallskip} 
CDFS    & $03^{\mathrm{h}} 32^{\mathrm{m}} 25^{\mathrm{s}}$ & $-27\degr 48' 50''$ &
        $223\fdg 6$ & $-54\fdg 5$ & 0.01 \\ 
A 901   & $09^{\mathrm{h}} 56^{\mathrm{m}} 17^{\mathrm{s}}$ & $-10\degr 01' 25''$ & 
        $248\fdg 0$ & $+33\fdg 6$ & 0.06 \\ 
S 11    & $11^{\mathrm{h}} 42^{\mathrm{m}} 58^{\mathrm{s}}$ & $-01\degr 42' 50''$ & 
        $270\fdg 5$ & $+56\fdg 8$ & 0.02 \\
\noalign{\smallskip} \hline
\end{tabular}
\end{table}

The COMBO-17 project (``Classifying Objects by Medium-Band Observations in 
17 Filters'') was designed to provide a sample of $\sim$50,000 galaxies and 
several hundred AGN with precise photometric redshifts and spectral energy
distributions (SEDs). As shown below, the filter set provides a redshift 
accuracy of $\sigma_z \approx 0.03$ for quasars \citep[and similarly for 
galaxies; cf.][]{Wol03}, smoothing the unknown true redshift distribution 
of the sample only very mildly and certainly allowing the derivation of 
luminosity functions.

The first step of the COMBO-17 data analysis was to convert the 
photometric observations into a very low resolution `fuzzy spectrum',
allowing for simultaneous spectral classification into stars, galaxies 
and QSOs, as well as for accurate redshift and SED estimation for the latter
two categories. The full survey catalogue will contain about 50,000 objects 
with classifications and redshifts covering a solid angle of 
1.0~$\mbox{deg}^2$. This {\it fuzzy spectroscopy} 
consciously compromises on redshift accuracy in order to obtain large 
samples of quasars with a reasonable observational effort.

While the photometric redshift technique has already been applied to galaxy 
samples about 40 years ago \citep{Baum62,But83}, we have modified and improved
the approach by increasing the number of filters and narrowing their bandwidth 
to obtain better spectral resolution and more spectral bins. This way,
COMBO-17 provides identifications and reasonably accurate redshifts
not only for galaxies but also for quasars, a novelty pioneered in 
CADIS \citep{Wol99} and more recently applied in the SDSS \citep{Ric01} 
and to observations with superconducting tunnel junctions \citep{dBr02}.

All observations presented here were obtained with the Wide Field Imager 
\citep[WFI,][]{WFI1} at the MPG/ESO 2.2-m telescope on La Silla, Chile. The 
WFI provides a field of view of $34\arcmin \times 33\arcmin$ on a CCD mosaic 
consisting of eight 2k $\times$ 4k CCDs with a scale of $0\farcs238$/pixel. 
The total observing time for COMBO-17 was about $\sim$175~ksec per field, 
including a $\sim$20~ksec exposure in the R-band with seeing below 0\farcs8. 

Observations and data analysis have been completed on three fields including
the Chandra Deep Field South (CDFS), covering an area of 0.78~$\mbox{deg}^2$ 
(Tab.~\ref{fields}). The deep $R$-band images have 5-$\sigma$ point source 
limits of $R\approx 26$ and provide the highest signal-to-noise ratio for 
object detection and position measurement among all data in the survey, 
except for L-stars and quasars at $z>5$, which can be extremely faint in R.

Using SExtractor \citep{BA96}, we obtained a catalogue of $\sim$ 200,000 
objects with positions and the SExtractor photometry MAG--AUTO \citep[for 
details on the construction of the catalogue see][]{Wol01}. This $R$-band 
selected source catalogue was then used to obtain SEDs across 17 passbands 
with a photometric technique specifically tailored for 
measuring colour indices with high signal-to-noise and as little as possible
interference by changing observing conditions. To this end, we projected 
the object coordinates into the frames of reference of each single exposure 
and measured the object fluxes at the given locations using with the 
well-established MPIAPHOT approach of a seeing-adaptive weighted 
aperture \citep{RM91}.  

The COMBO-17 photometric calibration is based on a system of faint standard 
stars in the COMBO-17 fields, which we tied to spectrophotometric standard 
stars during photometric nights. Our standards were selected from the 
Hamburg/ESO survey database \citep{WC00} of digital objective prism spectra. 
By having standard stars within each survey 
exposure, we were independent from photometric conditions for imaging. 
More details of the data reduction will be provided in a 
forthcoming technical survey paper (Wolf et al., in prep). 
In the present paper, all magnitudes are quoted with reference 
to Vega as a zero point.

\begin{figure}
\centering
\includegraphics[clip,angle=270,width=\hsize]{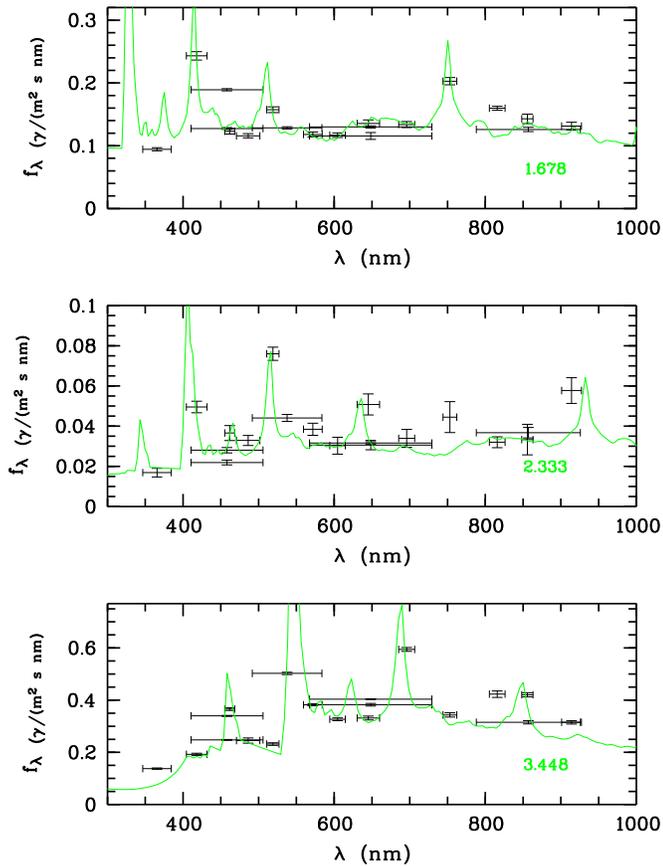}
\caption{Filter spectra of example quasars: The three panels show 
quasars at different redshifts across the range addressed
in this paper. The filter spectra are plotted with horizontal bars
resembling the filter width and vertical bars for 1$\sigma$ flux errors.
Due to variability repeated observations in B and R filters can show
different flux levels for the same object. Variability is compensated 
for the construction of SEDs using R-band images which are available 
for every observing run. The best-fitting template spectra at indicated 
redshifts are plotted as grey lines.}
\label{examples}
\end{figure}

\begin{figure}
\centering
\includegraphics[clip,angle=270,width=\hsize]{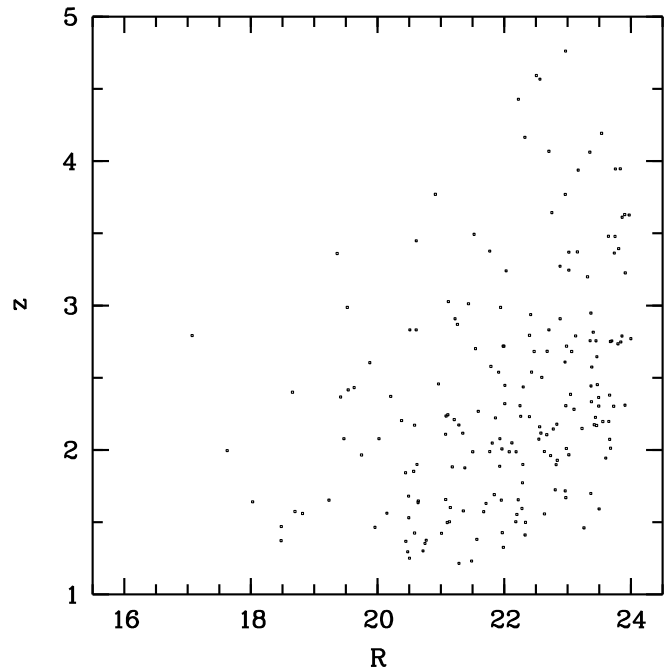}
\caption{The present sample of 192 quasars: 
Distribution of redshifts over observed R-band magnitude. Notice
that the sample has been truncated at $z=1.2$ and $R=24$.}
\label{qsosample}
\end{figure}

\section{The quasar catalogue}

\subsection{Sample definition}

The quasar sample is extracted from the full survey catalogue purely on 
the basis of spectral information, thus deliberately ignoring morphological 
evidence. We believe that the classification of the `fuzzy spectra' based on
17 filters, outlined below, allows to clearly differentiate 
between stars, galaxies and quasars, providing a safer separation 
between the object classes than morphological criteria. This is particularly
important in the context of our interest in low-luminosity AGN which may
well appear extended on a deep $R$-band image. Our purely spectroscopic
classification approach ensures that the sample will not be heavily biased
against such objects.

The sample finally used for all analyses in this paper is defined by limits 
in magnitude and redshift. Objects are selected to have a magnitude of $R>17$ 
to avoid the saturation regime in the individual frames, and $R<24$ which is
where the completeness of the quasar identification has dropped below 30\%
(see Sect.\ \ref{completeness}).  Since quasars show brightness variations, 
the sample selection will depend on the epoch of observation at the faint
end. Here, we have used our R-band photometry from January 2000.

The sample is further limited to redshifts of $z>1.2$, since host galaxies 
may contribute significantly to the spectra at lower redshifts, where the 
4000~$\mathrm{\AA}$-break is still contained within the filterset. Our 
templates currently contain only pure quasar and pure galaxy spectra, but no 
mixed templates with contributions from both. Therefore, identification and 
redshift estimation of low-luminosity quasars at $z<1.2$ is not straightforward
at this point, and complicated completeness issues arise in the low-$z$ domain.

In order to ensure minimum contamination from non-active galaxies, we have 
set our probability threshold for an object to be classified as quasars 
quite high. As a result, we have eliminated many trustworthy low-luminosity 
AGN from the sample, but keep a well-controlled selection of 
higher-luminosity QSOs. We have chosen this conservative approach 
because we have not yet obtained spectroscopic confirmation redshifts 
for a sufficiently large sample to understand the selection function of
low-luminosity Seyfert galaxies well enough.

The catalogue resulting from this selection contains 192 quasars between 
$z=1.2$ and $z=4.8$, with a median redshift of $\langle z\rangle=2.23$.
Examples of quasar filter spectra are shown in Fig.~\ref{examples}, and the 
Hubble diagram for the full sample is presented in Fig.~\ref{qsosample}.

\subsection{Classification and redshift estimation}

The photometric measurements from 17 filters provide low-resolution 
spectra for each object which are analysed by a statistical technique for 
classification and redshift estimation based on spectral template matching 
\citep[for details see][]{WMR01}. Meanwhile, we have improved the template 
library for quasars by deriving it from the recent SDSS QSO template 
spectrum \citep{vdB01}, rather than from the earlier used 
pure emission line contour by Francis et al. (1991). 
This leads to a better detection and redshift estimation of 
quasars at $z<2.5$ where the `little blue bump' may render the spectral shape 
between the prominent emission lines as quite different from the power-law 
which we assumed for the quasar templates previously. Since the CADIS work
\citep{Wol99}, we have also learned that photometric redshifts for quasars 
are strongly influenced by flux variability. 

Since the multi-colour observations were collected over a period of two years,
and quasars as well as some stars show variability, it was necessary to correct
for variability when constructing the 17-filter spectra for classification. 
For this purpose we used the $R$-band observations which are available for
each observing epoch, allowing us to completely map the variability at least 
in this band. When constructing the SEDs of variable objects, 
we related the measurements of observed photometric band
to the $R$-band magnitude obtained in the same observing run. 
As a result, the SEDs are not distorted by long-term magnitude changes. 
This variability correction works well only as long as flux variations are not
accompanied by major changes in spectral shape. Furthermore, we
also cannot account for short-term variability on a time scale of days,
as within each observing run we have obtained only one $R$-band image.

\begin{figure*}
\sidecaption
\includegraphics[clip,width=12cm]{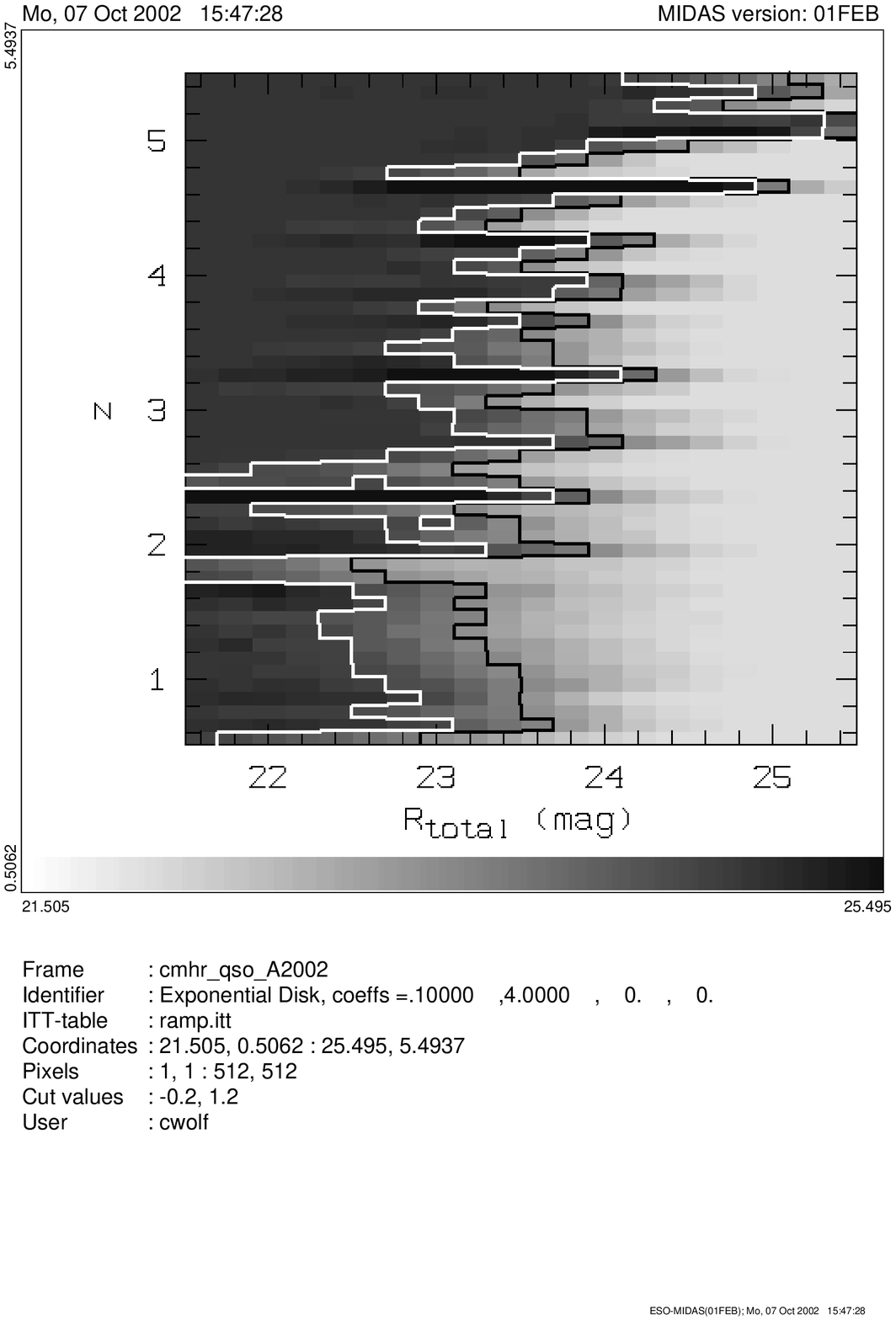}
\caption{Completeness map for quasar selection and redshift estimation: 
Grey-scale and contour maps demonstrating how the fraction of quasars 
having successful redshift measurements depends on magnitude and redshift. 
Completeness levels are shown as a greyscale from 0\,\% (light grey) to 
120\,\% (black). Contour lines are drawn for 90\,\% (white) and 
50\,\% completeness (black). Values above 100\,\% occur when 
redshift aliasing creates local overdensities in the estimated 
$z$-distribution based on a flat underlying simulated distribution. 
See Sect.~\ref{completeness} for a more detailed discussion of 
the completeness correction.}
\label{cmap}
\end{figure*}

\subsection{Completeness correction}\label{completeness}

The subsequent analysis includes only objects with successful $z$ estimates. 
It is therefore critical to understand for which quasars the data permit 
such a classification. Given the photometric properties of the survey,
we can produce mock catalogues of stars, galaxies and quasars
with realistic photometric errors, and investigate the classification 
performance as a function of object type, redshift and magnitude. The
17-filter-spectra in the mock catalogues are constructed from the 
library templates, using the empirically derived observational 
error distributions. With this approach, the completeness of the 
classification and redshift 
estimation can be derived from Monte-Carlo simulations \citep[see][]{WMR01}. 
We implicitly assume that the spectral templates truly resemble observed 
objects, and if that is not the case, our simulation will be too optimistic.

In fact, we can test whether our completeness maps are realistic,
given that within our fields and selection limits 12 broad-line AGN are
known from spectroscopic observations in the CDFS (Hasinger 2002, priv.
comm.; Szokoly et al, in prep.). The map predicts that 10 out of 12
objects should be identified, while in fact we recover 8. Two objects
are missing from our sample although they lie in regions of high expected
completeness: A QSO with broad absorption lines (BALs) at $z=3.6$ and a
Sy-1 galaxy at $z=1.2$. The BAL QSO was misclassified as a star because of
its unusually star-like colours. The Sy-1 galaxy resides just at our low
redshift limit which we adopted to avoid incompleteness arising from the
pure AGN spectra getting contaminated by host galaxy contributions.
We conclude that on the whole our map is realistic, but that (a) rare QSOs
with unusual colours could still escape our attention, and (b) right at the
low-redshift limit the level of completeness might be reduced compared
to our maps.

The product of the simulations for quasars is a completeness map 
(selection function), providing a formal probability  $C(R,z,SED)$ 
that a quasar of given intrinsic properties is recovered, 
in bins of observed $R$-band magnitude and redshift.
Our quasar template library contains a range of emission line strengths and
a range in spectral indices or continuum colours. Effective spectral indices
depend on redshift as the quasar continuum is not a pure power law. However,
for a quasar at $z=2.0$ the range of template $B-I$ colours runs from about 
$+0.35$ to $+1.75$, corresponding roughly to power law indices of $\alpha =
-1.66\ldots +0.4$. After we recognized that the simulated completeness
depends very little on the spectral index, we collapsed the completeness 
function into $C(R,z)$ removing the explicit SED dependence
(see Fig.~\ref{cmap}).

At most redshifts, our classification algorithm is more than 90\% complete 
for quasars with $R\la23$. The contour line for 50\% completeness ranges 
around $R\sim23.5$. The redshift dependence of the completeness shows 
conspicuous oscillations above redshift 2, which are caused by the strong 
signature of the Lyman-$\alpha$ emission line migrating through the filter set 
and alternating between visibility in a medium-band filter and invisibility 
when it falls between two neighboring medium-band filters. Whenever the line 
is visible quasars can be identified to fainter levels due to its contrast.

At certain redshifts, the selection function reaches values above 100\,\%.
This occurs when redshift aliasing creates local overdensities 
in the estimated $z$-distribution based on a flat underlying simulated 
distribution. Such `overcomplete' regions are always accompanied by 
adjacent `undercomplete' zones, so that the total number of objects 
is conserved.

\begin{figure*}
\centering
\includegraphics[clip,angle=270,width=16cm]{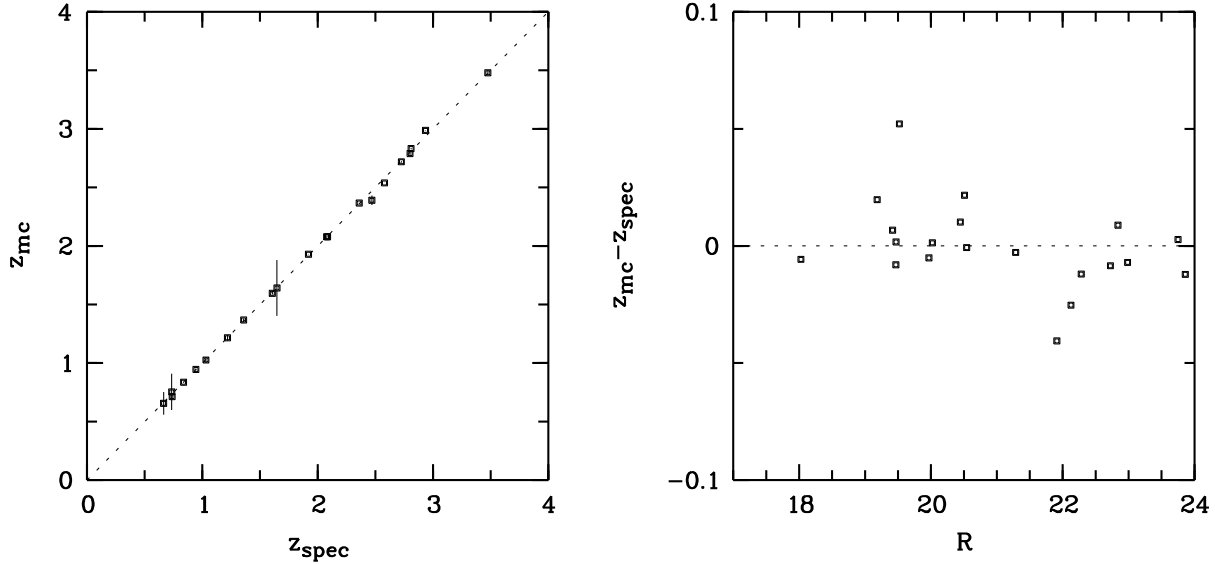}
\caption{Spectroscopic vs. multi-colour redshifts: 22 QSOs from the
COMBO-17 sample at $R<24$ have spectroscopic identifications. 21 of them
were found to be QSOs at redshifts within $\pm 0.05$ of the multi-colour
estimate, one of them is a White Dwarf/M dwarf-pair estimated at $z<1$.
\label{zz_comp}}
\end{figure*}

\subsection{Influence of limited redshift accuracy}

For the analysis it is crucial to check to which extent the limited 
redshift accuracy  of $\sigma_z\approx 0.03$ (as determined from the 
simulations) could affect inferences about luminosity functions. 
Two major aspects need to be explored, {\it redshift aliasing} 
and {\it catastrophic mistakes}, which are both irrelevant for interpreting 
well-exposed data from an aperture spectrograph, 
but could both play a role in our case: 
\begin{enumerate}
\item 
Aliasing results from structures which are finer than the redshift resolution,
violating the sampling theorem; it appears as fake structure on a scale
that is typically slightly larger than the resolution. This effect is
doubtlessly present in our data, as is also visible from the $z$-dependent
oscillations in Fig.\ \ref{cmap}. In this paper, we avoid dealing with the 
problem altogether by only considering redshift bins of 
typically $\Delta z = 0.6$, an order of magnitude larger 
than the resolution limit.

\item
Catastrophic mistakes occur in certain regions of colour space where different 
interpretations can be assigned to the same colour vector and probabilistic
assumptions are used to make a final redshift assignment. Obviously, a number
of cases could lead to the assignment of a wrong redshift, but their effect on 
luminosity functions should be relatively small unless a major fraction of 
objects were affected. The only real impact would be in a situation in which 
many low-redshift objects of medium luminosity were wrongly assumed to reside 
at high redshift, boosting the abundance of the rare luminous quasars. 
We see no evidence for any such effect to be important.

\end{enumerate}

A large-scale performance check of our spectrophotometric redshifts 
based on a substantial sample of spectroscopic redshifts 
is still pending. 
However, we have been able to test our classifications and redshifts
against two samples of albeit limited sizes that have spectroscopic data. 
In the S11 field, the COMBO-17 sample overlaps at its bright end 
with the `2QZ' quasar survey \citep{Cro01}, and in the CDFS we 
found that a number of spectroscopically identified 
X-ray sources coincide with faint COMBO-17 AGN
(Hasinger 2002, priv.comm.; Szokoly et al, in prep.).
Fig.~\ref{zz_comp} shows a comparison of spectroscopic and multi-colour 
redshifts for all 22 objects from the COMBO-17 QSO sample with $R<24$ 
for which spectra exist. Only one of these 22 objects, 
classified as a QSO at $z\sim0.6$ by COMBO-17 
(i.e.\ outside the redshift range considered in this paper), 
was actually a misclassification; a 2QZ spectrum identified 
this object correctly as a White Dwarf/M dwarf-pair. 
The other 21 objects were found to be QSOs, and their redshifts 
agree all within $\pm0.05$. This comparison suggests 
that the accuracy of COMBO-17 redshifts is certainly no worse 
than the assumed $\sigma_z = 0.03$, and the rate of catastrophic 
mistakes is probably well below 10\,\%. 

The CDFS contains several further faint Seyfert galaxies, 
particularly at $z<1$, most of which were recently identified 
as optical counterparts to faint X-ray sources.
These are very challenging objects for the multi-colour redshift estimation, 
as their host galaxies will contribute significantly to their overall SED. 
Clearly, more work is needed to tackle these objects, which are therefore
explicitly excluded from the scope of this paper.

\begin{figure}
\includegraphics[clip,width=\hsize]{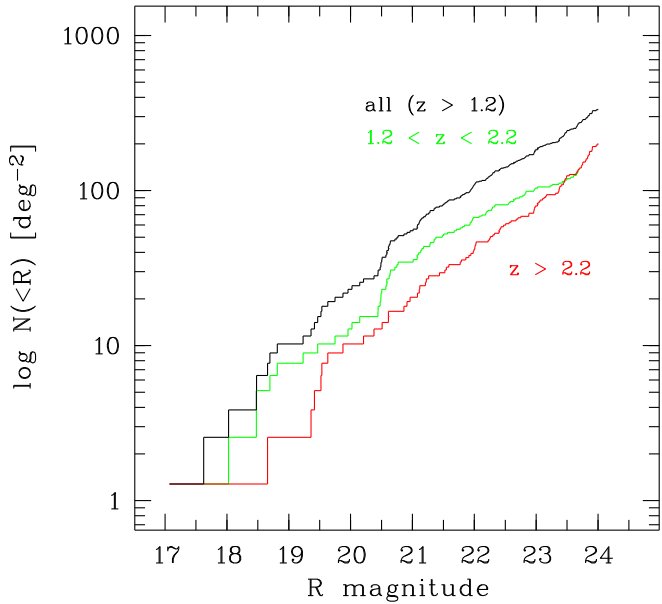}
\caption[]{Cumulative surface density of AGN as a function of
$R$ band magnitude, corrected for Galactic extinction.}
\label{fig:surfdens}
\end{figure}

\begin{table}
\caption[]{Tabulated AGN number counts as a function of $R$ band 
magnitude. The 2nd and 4th column give the actual numbers of
objects in the current COMBO-17 sample (within 0.78~deg$^2$), 
while the 3rd and 5th column give incompleteness-corrected 
surface densities per deg$^2$.}
\label{tab:surfdens}
\begin{tabular}{r@{\hspace*{2.5em}}rr@{\hspace*{2.5em}}rr}
\hline \hline
\noalign{\smallskip}
        & \multicolumn{2}{c}{$z > 1.2$\hspace*{2.5em}} &
          \multicolumn{2}{c}{$z > 2.2$\hspace*{0.5em}} \\
$R$ & \multicolumn{1}{c}{$n$} & \multicolumn{1}{c}{${\cal N}(<R)$\hspace*{1.8em}} &
      \multicolumn{1}{c}{$n$} & \multicolumn{1}{c}{${\cal N}(<R)$} \\
\noalign{\smallskip}\hline\noalign{\smallskip}
24.0 & 192 & 336.5 & 101 & 201.7 \\
23.0 & 139 & 183.2 &  60 &  79.5 \\
22.0 &  86 & 107.7 &  33 &  41.8 \\
21.0 &  43 &  55.1 &  16 &  20.5 \\
20.0 &  18 &  23.1 &   8 &  10.3 \\
19.0 &   8 &  10.3 &   2 &   2.6 \\
18.0 &   2 &   2.6 &   1 &   1.3 \\
\noalign{\smallskip}\hline\noalign{\smallskip}\\
\end{tabular}
\end{table}

\section{Number counts}

From the AGN sample and the completeness map described above
it is straightforward to compute surface densities as a 
function of apparent magnitude, using the relation
\begin{equation}
{\cal N}(<R) \:=\: \sum_i \frac{1}{A_i} 
             \:=\: \frac{1}{A_0}\,\sum_i \frac{1}{C (R_i,z_i)} 
\end{equation}
where the summation runs over all AGN brighter than $R$.
$A_0$ is the formal total survey area of 0.78~deg$^2$;
multiplying this with the value of the completeness map 
$C(R_i,z_i)$ for object $i$ yields the estimated 
`effective survey area' $A_i$ for that particular source.
In other words, the decreased probability of detecting 
very faint AGN, particularly relevant for certain redshift 
ranges, is simply interpreted as being equivalent to a 
reduced survey area for such objects.

Results are shown in Fig.\ \ref{fig:surfdens}
and listed in Table \ref{tab:surfdens}. Besides
the full sample of all $z>1.2$ AGN, we also show the 
number counts for subsamples split at $z < 2.2$ and
$z > 2.2$, the typical high-redshift limit of UV excess surveys.
The two subsamples show very different trends towards
the faint limit: While the low-$z$ objects dominate at 
brighter magnitudes, their surface density increases only 
slowly with decreasing flux level. On the other hand, 
the cumulative number of AGN with $z>2.2$ is a strong
function of magnitude, and quite well described 
by a power law ${\cal N}(<R) \propto R^{-0.31}$, 
corresponding to a power law index of $-1.75$ for the
differential number-flux relation. Towards the faint end
it even seems as if the slope might be increasing, but this
is not formally significant. Nevertheless, for $R > 23$
it is clear that high-redshift AGN with $z>2.2$ become 
the dominant population.

\begin{figure}
\includegraphics[clip,width=\hsize]{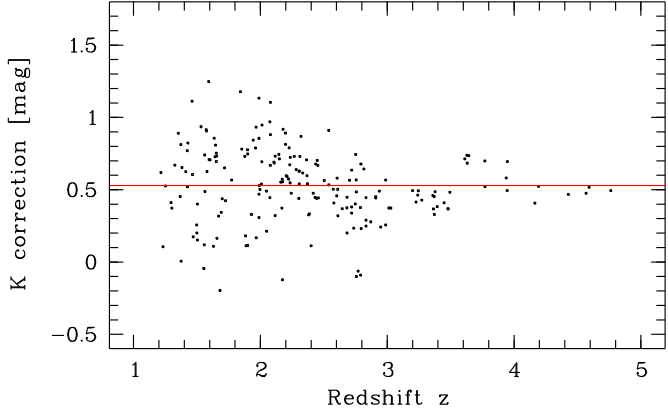}
\caption[]{Derivation of the internal K correction used in this paper: Each
  point corresponds to one individual AGN measurement based on the SED fits
  within COMBO-17. The horizontal line is a linear fit, allowing to predict
  luminosities at 145~nm with an accuracy of 0.24~mag.}
\label{fig:kcorr}
\end{figure}

\begin{figure}
\includegraphics[clip,width=\hsize]{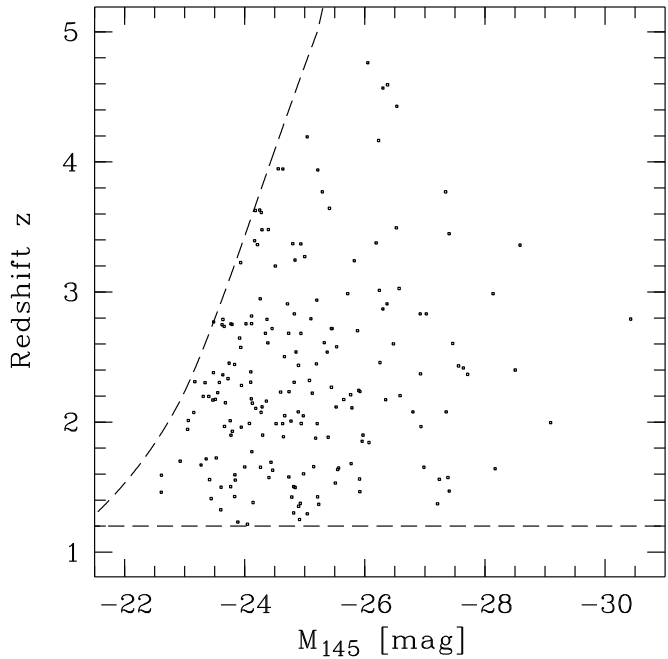}
\caption[]{Distribution of the input sample over absolute magnitudes
and redshifts: The dashed lines indicate the imposed sample limits of $z>1.2$ 
and $R<24$.}
\label{fig:zmabs}
\end{figure}

\section{Estimation of rest frame luminosities}

For each object in the sample we have individual SEDs from the 
17-filter spectrophotometry at our disposal.
In order to derive absolute magnitudes for the subsequent analysis, 
we decided to tie all luminosities to the UV continuum level at 
$\lambda_{\mathrm{rest}} = 145$~nm.
This was realised by integrating the best-fitting redshifted template 
over a synthetic narrow rectangular passband at 143~nm--147~nm, 
thus avoiding the \ion{Si}{iv}/\ion{O}{iv} emission line at 140~nm.
This procedure enabled us to directly measure rest-frame luminosities
over a redshift range $1.4 \la z \la 5$ without any need for extrapolation.

While such individually determined luminosities may suffer less from biases
arising in uncertain assumptions on AGN spectral energy distributions than
some previous samples, they considerably complicate the statistical 
exploitation. Although most of our objects are detected in all or nearly all
of the 17 filter bands, the sample is presently \emph{defined} only in the 
$R$-band: (1) The flux limit is homogeneously truncated at $R<24$. (2) The
completeness map has the $R$ band magnitude as one of its two independent
parameters. It is also assumed in the algorithms used for 
luminosity function estimation that the survey is defined by 
just one flux limit. We have therefore explored whether it is possible 
to estimate absolute UV magnitudes $M_{145}$ from the measured $R$ band
flux, using a modified but basically traditional $K$ correction approach.

For this purpose we computed for each quasar its absolute magnitude
directly from the distance modulus, $M_0 = R - (m-M)$, equivalent to
assuming a $K$ correction for a power-law spectrum with slope $\alpha = -1$.
We then plotted the differences $\Delta M$ between $M_0$ and $M_{145}$ (the 
spectrophotometric estimates) against source redshift; this is shown as the 
distribution of points in Fig.\ \ref{fig:kcorr}. 
This plot shows that at least at redshifts $z\ga 2$ there is a 
well-defined relation between $z$ and $\Delta M$. When fitting a constant 
line through these points, as shown in Fig.\ \ref{fig:kcorr}, we get an 
overall rms scatter of 0.24~mag (0.13~mag for $z>2.2$ and 0.26~mag for 
$z < 2.2$), which can only insignificantly be improved by 
fitting e.g. a low-order polynomial. This line defines then 
an empirical correction relation $K(z)$, so that for the redshift range 
of interest in this paper we can use $R$ band magnitudes 
to estimate $M_{145}$ with an accuracy of 0.24~mag, 
even without using any SED information.

Most earlier studies of the optical AGN luminosity function, especially 
those focussing on lower redshifts, have expressed their results in
terms of blue magnitudes $M_B$. In order to facilitate
a statistical comparison, we obtained a crude estimate for the 
offset $M_B - M_{145}$ by repeating the above procedure for 
the rest-frame $B$ band. Note that for all $z>1.2$ objects this
involves a good deal of extrapolation, and the diagram corresponding
to Fig.\ \ref{fig:kcorr} shows more than 0.6~mag of scatter around
the mean trend. Nevertheless, the offset of 1.75~mag thus determined
is actually identical (to 0.01~mag) to the result when the same quantity
is measured in the mean quasar energy distribution of Elvis et al. (1994).
In conclusion this means that all quoted absolute magnitudes $M_{145}$
can be converted into $M_B$ using the relation $M_B = M_{145} + 1.75$.

In the subsequent analysis we use two sets of cosmological parameters,
an Einstein-de Sitter universe with $H_0=50$~km~s$^{-1}$~Mpc$^{-1}$,
$\Omega_m = 1.0$ and $\Omega_\Lambda=0$, and a flat `concordance model'
with $H_0=65$~km~s$^{-1}$~Mpc$^{-1}$, $\Omega_m = 0.3$ and 
$\Omega_\Lambda=0.7$.
While the former is now physically almost obsolete, it still has some
relevance as a `yardstick' model, as most earlier studies of AGN evolution
were expressed preferentially in these terms. Actually we found that 
the results change very little when switching between the two models,
except for small shifts in both axes,
and our displayed results always refer to the `concordance universe'
unless explicitly stated otherwise. Figure \ref{fig:zmabs} shows the
distribution of absolute magnitudes vs.\ redshift for the entire
AGN sample. In terms of the conventional (if arbitrary) distinction
between high-luminosity quasars and low-luminosity Seyferts
around $M_B \simeq -23$, our sample covers just the region close
to this dividing line. It is therefore the first optically 
selected AGN sample of substantial size to probe this important
regime of nuclear activity in galaxies. It also matches very
well the luminosity range covered by low-$z$ AGN surveys that
provide a $z\simeq 0$ reference \citep[e.g.,][]{Koe97}.

\begin{figure*}
\includegraphics[clip,width=16cm]{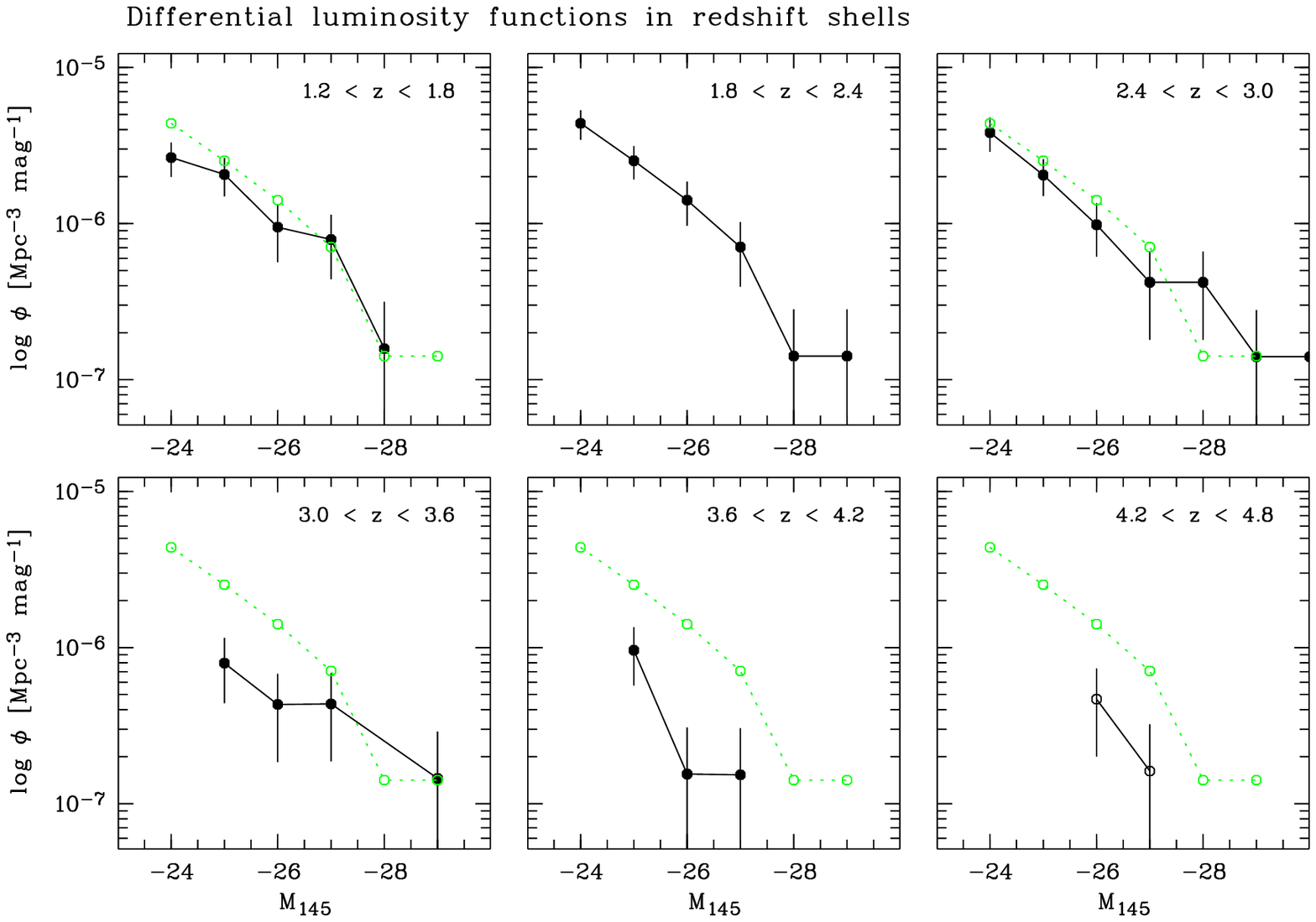} \\[2ex]
\includegraphics[clip,width=16cm]{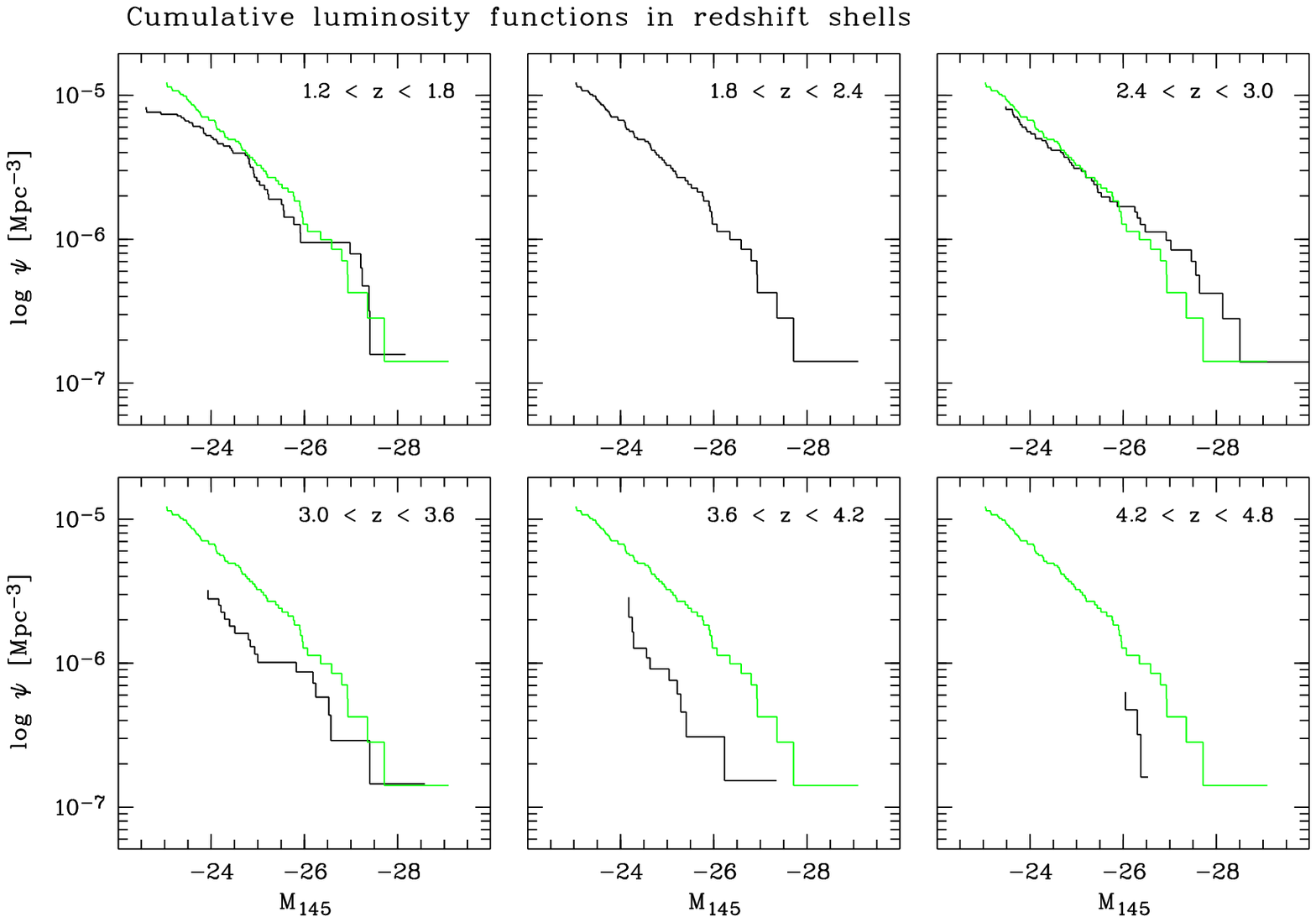}
\caption[]{(a) Binned differential luminosity functions for 
six non-overlapping redshift shells. Only luminosity bins completely
covered by the sample are shown. In addition to the data
with Poissonian error bars, each panel features the $1.8 < z < 2.4$
luminosity function as reference.
(b) Cumulative luminosity functions for the same redshift shells.}
\label{fig:lf}
\end{figure*}

\begin{table*}
\caption[]{Tabulated binned AGN luminosity function.}
\label{tab:binlf}
\begin{tabular}{rr@{\hspace*{2.5em}}rr@{\hspace*{2.5em}}rr@{\hspace*{2.5em}}rr@{\hspace*{2.5em}}rr@{\hspace*{2.5em}}rr@{\hspace*{2.5em}}rr}
\hline \hline
\noalign{\smallskip}
        & & \multicolumn{2}{c}{1.2--1.8\hspace*{2.5em}} &
            \multicolumn{2}{c}{1.8--2.4\hspace*{2.5em}} & 
            \multicolumn{2}{c}{2.4--3.0\hspace*{2.5em}} &
            \multicolumn{2}{c}{3.0--3.6\hspace*{2.5em}} &
            \multicolumn{2}{c}{3.6--4.2\hspace*{2.5em}} &
            \multicolumn{2}{c}{4.2--4.8\hspace*{0.5em}} \\
$M_{145}$ & $M_B$ & 
            $n$ & $\log\phi$ & $n$ & $\log\phi$ & $n$ & $\log\phi$ &
            $n$ & $\log\phi$ & $n$ & $\log\phi$ & $n$ & $\log\phi$ \\
\noalign{\smallskip}\hline\noalign{\smallskip}
$-$23.0 & $-$21.2 &  7 & $-$5.78 &  9 & $-$5.53 &  0 &         & 0 &         & 0 &         & 0 \\
$-$24.0 & $-$22.2 & 16 & $-$5.58 & 22 & $-$5.36 & 17 & $-$5.42 & 5 & $-$5.85 & 3 & $-$6.02 & 0 \\
$-$25.0 & $-$23.2 & 13 & $-$5.69 & 17 & $-$5.60 & 14 & $-$5.69 & 5 & $-$6.10 & 6 & $-$6.02 & 0 \\
$-$26.0 & $-$24.2 &  6 & $-$6.02 & 10 & $-$5.85 &  7 & $-$6.01 & 3 & $-$6.36 & 1 & $-$6.81 & 3 & $-$6.33 \\
$-$27.0 & $-$25.2 &  5 & $-$6.10 &  5 & $-$6.15 &  3 & $-$6.38 & 3 & $-$6.36 & 1 & $-$6.81 & 1 & $-$6.79 \\
$-$28.0 & $-$26.2 &  1 & $-$6.80 &  1 & $-$6.85 &  3 & $-$6.38 & 0 &         & 0 &         & 0 \\
\noalign{\smallskip}\hline\noalign{\smallskip}
\end{tabular}
\end{table*}

\section{Evolution of the AGN luminosity function}

\subsection{Non-parametric estimates}

\subsubsection{Method of computation}

We employ the usual $1/V_\mathrm{max}$ estimator \cite{Sch68}
to give the space density contributions of individual objects.
Luminosity functions are then readily obtained by forming 
the appropriate sums:
Denoting the luminosity-binned differential LF as $\phi(M)$
and the cumulative LF as $\psi(M)$ we have 
\begin{eqnarray}
\phi(M,z) & = & \sum_{M-\Delta M}^{M+\Delta M} \:
                \sum_{z-\Delta z}^{z+\Delta z} \frac{1}{V_i}\:, \\
\psi(M,z) & = & \sum_{-\infty}^{M}
                \sum_{z-\Delta z}^{z+\Delta z} \frac{1}{V_i} \nonumber
\end{eqnarray}
where each value of $V_i$ is the total comoving volume within which 
object $i$ would still be included in the sample. 
The value of $V_i$ follows from integrating the survey selection 
function $C(R,z)$, introduced in Sect.~\ref{completeness}, over
all relevant redshifts:
\begin{equation}
 V_i (R,z) = A_0 \int_z^{z_\mathrm{max}}
                {C (R,z) \frac{dV}{dz}dz}
\end{equation}
\citep[cf.][]{Wis98} where $A_0$ is again the formal total 
survey area, $A_0 = 0.78$~deg$^2$ in the present case.
The assigned error bars are based purely on Poissonian shot noise due to 
the limited AGN counts within each bin, i.e.\ we ignored errors 
in magnitude or redshift estimation.
The error bars for a given sum are then given by 
\begin{equation}
 \sigma_\phi = \sqrt{\sum_i 1/V_i^2 (M,z)}.
\end{equation}
It is well known that binned estimates of luminosity functions
suffer from a number of biases. In particular, two effects
are worth being recalled, both of which can be seen as two
different variants of Malmquist bias and lead to the observed
LF appearing flatter than the intrinsic one:
\begin{itemize}
\item Binning in magnitude and a steep LF create a positive bias
in the derived space densities, due to a shift in effective bin centres
towards lower luminosities.
\item Differential evolution within redshift shells leads to an
overestimation of space densities where the LF is steep.
\end{itemize}
While the first effect is absent in the cumulative LF representation, 
the latter is unavoidable as long as no model assumptions for the
evolution of space densities are made. 
It should thus be kept in mind that the nonparametric LFs discussed
in the following section do not necessarily show the data `as they
are', but just form the most conventional way of presenting 
luminosity function data.

\subsubsection{Results}

Inside the redshift interval $1.2 < z < 4.8$ covered by the COMBO-17 
sample we defined six redshift shells of equal sizes to trace the evolution
of the QLF. Figure \ref{fig:lf} gives a synopsis of the results;
for the benefit of interested readers who wish to use our data
for their own computations we present the tabulated binned LFs 
in Table~\ref{tab:binlf}. This table shows also how many objects
contributed to each bin. The numbers given in this table show 
that the differences between adjacent redshift shells are not huge,
i.e.\ evolution is relatively moderate over this redshift range. 
We therefore have deviated from the usual practice to plot all LFs 
into one frame, as any real trend would be very hard to discern 
in such a diagram.
Instead, we created individual subpanels for each redshift shell and
plot each luminosity function separately, but together with the 
$1.8 < z < 2.4$ LF taken as a reference. This visualisation shows 
that the measured values of the LF are below the corresponding 
reference values in nearly all data points. In other words, we detect
an unambiguous maximum in the comoving AGN space density
near these redshifts, and a significant drop towards both
lower and higher $z$. The existence of such a maximum becomes even 
clearer in Fig.\ \ref{fig:spacedens} where integrated space density 
at given lower bound in luminosity (taken directly from the cumulative 
LFs in each redshift shell) is plotted directly against $z$.
We return to a more detailed discussion of the evolution of 
space densities and LF shape properties below.

\subsection{Parametric analysis of the luminosity function}

\subsubsection{Ansatz}

Within certain bounds and accuracy limits, an observed 
luminosity function can usually be described quite well
by some simple analytic expression, allowing one to 
compress the results into a few well-determined numbers,
with the positive side effect that the above mentioned 
Malmquist-type biases due to data binning can be avoided 
by obtaining luminosity function and evolution parameters simultaneously 
from an observed sample. This was first demonstrated by Marshall et al. 
(1983) for a very simple power law model of the evolving QLF. 
Since the parametric forms employed for the analysis in this paper
are somewhat non-standard, we use the following paragraphs
to spell out the adopted ansatz.

\paragraph{Shape of the luminosity function: }

The most common analytic description of the QLF is
a double power law, involving a bright-end slope $\gamma_1$,
a faint-end slope $\gamma_2$, and a smooth turnover
at a characteristic `break' luminosity $M^\star$. We have explored
this ansatz and decided not to use it for the present
sample, mainly because we fail to identify a well-defined
break luminosity in the data. Direct fitting of a double 
power law LF to any of the redshift shells subsamples shows
that $M^\star$ is a very ill-constrained quantity that often
takes a value outside the luminosity range covered by the sample; 
furthermore, even minor modifications in the subsample definition 
can cause major changes in $M^\star$.
Since the luminosity functions in Fig.\ \ref{fig:lf} undoubtedly
show \emph{some} indication of curvature, we decided to 
parametrise the LF, at given $z$, as a polynomial in $M_{145}$:
\begin{equation}
\log\phi \:=\: \sum_{i=0}^{n}\:A_{i}\,\mu^i\:.
\label{eq:lf_ansatz}
\end{equation}
where $\mu = M_{145}-M^\star$ and $M^\star$ is a constant
which can be chosen freely but may also contain the effects
of luminosity evolution (see below). 
In this representation, $A_0 = \log\phi_0$ is the space density
normalisation at $M_{145} = M^\star$, which in our case has
mostly been set to $M^\star = -25$
(roughly corresponding to $M_B = -23.25$). For $n=1$, the LF
is a single power law with slope $\gamma = -2.5 A_1 - 1$. 

\paragraph{Evolution parameters: }
The last decade has seen a sometimes hot debate over the question
whether `pure luminosity' or `pure density' evolution were more
appropriate. We have implemented both parametrisations as well
as mixed forms. As independent variable we use 
\begin{equation}
\zeta \:=\: \log \left (\frac{1+z}{1+z_{\mathrm{ref}}}\right ) 
\label{eq:zeta}
\end{equation}
which vanishes for $z = z_{\mathrm{ref}}$. For our analysis we
avoid extrapolating to redshifts outside the sample
definition interval and adopt $z_{\mathrm{ref}} = 2$. 
Pure density evolution is then expressed as polynomial 
expansion of the space density normalisation:
\begin{equation}
\log\phi_0 (z) \:=\: \sum_{j=0}^{m_D}\:C_{j}\,\zeta^j\:.
\label{eq:pde}
\end{equation}
For $m_D = 1$, the parameter $C_1$ is equal to the 
density evolution index $k_D$ used in several earlier analyses.

On the other hand, pure luminosity evolution involves a redefinition
of the LF parameter $M^\star$:
\begin{equation}
M^\star (z) \:=\: M^\star(\zeta = 0)\,+\,\sum_{j=1}^{m_L}\:B_{j}\,\zeta^j\:.
\label{eq:ple}
\end{equation}
(Note that unlike for a `broken power law', $M^\star(\zeta = 0)$ is here 
an arbitrary constant and not a free fit parameter.)
Again, the case of $m_L = 1$ corresponds to a standard
form already used by several previous authors, $M^\star(z) \propto (1+z)^{k_L}$.
In this case the luminosity evolution index is $k_L = 0.4 B_1$.

\paragraph{Fitting method: }
We followed a maximum likelihood approach
\citep{Mar83}
to search for optimal parameter combinations,
using a modified downhill-simplex minimisation scheme
\citep{Pre92}. 
Several goodness-of-fit tests were applied to check whether
the best-fit model was adequate.
These included one- and two-dimensional Kolmogorov-Smirnov 
tests for the distributions over the $(M,z)$ plane,
as well as a $\chi^2$ test that directly compares observed 
and predicted numbers of AGN in redshift/luminosity bins.
A fit was considered satisfactory only when it passed all
of the statistical tests with a probability for the validity
of the null hypothesis of $> 10$\,\%. This highly conservative
approach ensures that the missing distribution-free properties 
of the KS test \citep[see Wisotzki 1998 for discussion on][]{Lil68} 
can be neglected.

\begin{figure*}
\includegraphics[clip,width=17cm]{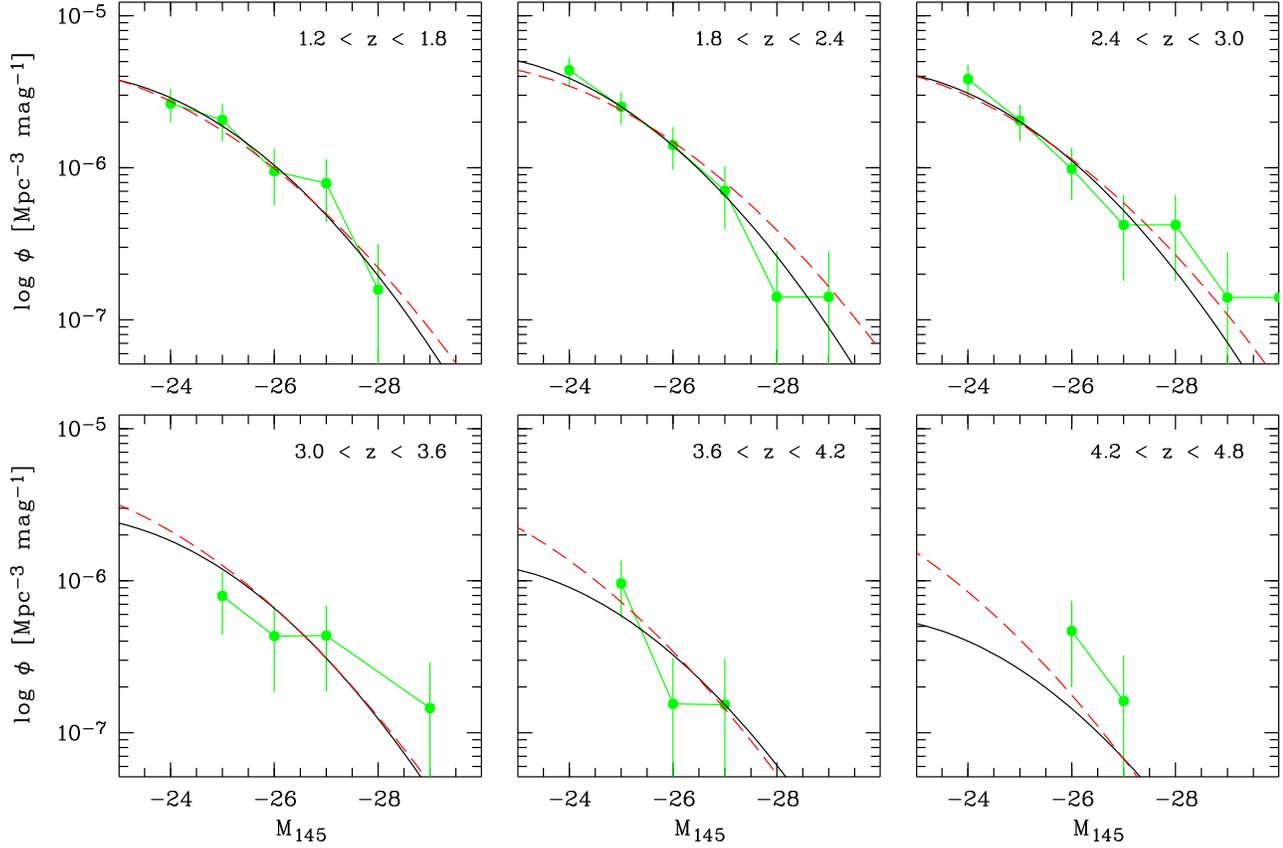}
\caption[]{Best-fit parametric representations of the evolving luminosity
function, plotted against the binned LF data from Fig.\ \ref{fig:lf}(a).
The density evolution model is shown by the solid, the luminosity evolution
model by the dashed lines.}
\label{fig:parametric}
\end{figure*}

\begin{figure*}
\includegraphics[clip,width=17cm]{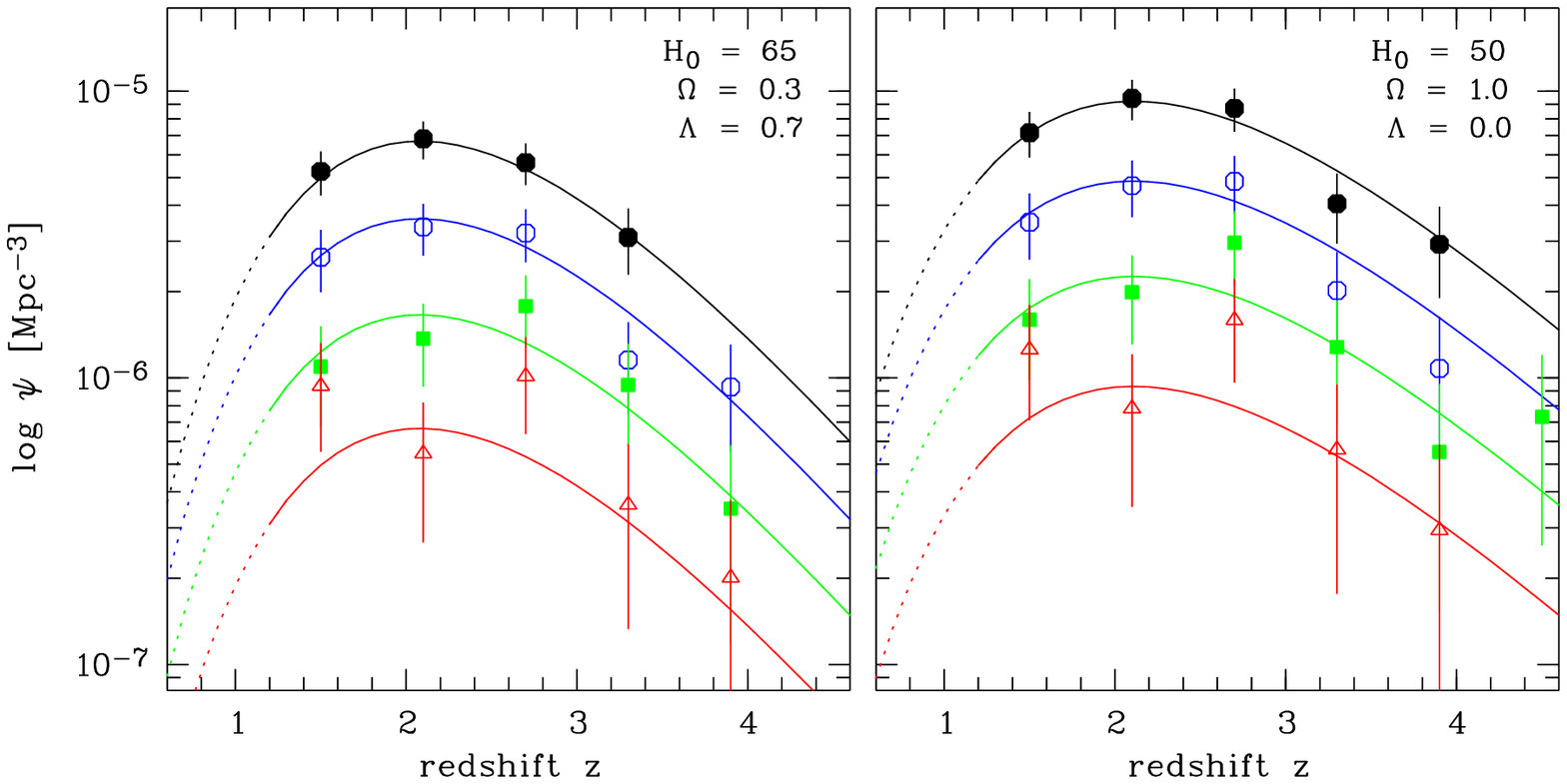}
\caption[]{Evolution of comoving AGN space density with redshift, for different
lower luminosity limits, and for two cosmological models. 
Filled circles: $M_{145} < -24$; open circles: $M_{145} < -25$; 
filled squares: $M_{145} < -26$; open triangles: $M_{145} < -27$.
The corresponding curves are integrated from the best-fit PDE models.
}
\label{fig:spacedens}
\end{figure*}

\subsubsection{Results}

Starting with the simplest possible luminosity function, we found that a 
single power law LF as an overall shape model had to be rejected as there 
is weak but significant curvature in the LF, especially for the lower 
redshift ranges, $z \la 2.4$. But already a 2nd order poynomial gives a 
statistically adequate description of the LF at all redshifts. Note that 
this form involves one free parameter less than the case of a double 
power law LF. 

{\bf  
\begin{table*}
\caption[]{Coefficients of the best-fit analytic models describing the
luminosity function and its evolution as either PDE or PLE. For details
on the notation see text.}
\label{tab:parameters}
\begin{tabular}{rrrrrrrrrr}
\hline \hline
\noalign{\smallskip}
Model & $A_0$ & $A_1$ & $A_2$ & $M^{\star}_{145}(0)$ & $C_1$ & $C_2$ & $B_1$ &
$B_2$ & $B_3$ \\
\noalign{\smallskip}\hline\noalign{\smallskip}
PDE & $-$5.600 & 0.2221 & $-$0.03536 & $-$25 & 0.3599 & $-$15.574 \\
PLE & $-$5.620 & 0.1845 & $-$0.02652 & $-$25 &           &        & 1.3455 &
$-$80.845 & 127.32 \\
\noalign{\smallskip}\hline
\end{tabular}
\end{table*}
}

As a first evolution mode we explored pure density evolution
(PDE). A global PDE fit over the entire redshift range 
$1.2 < z < 4.8$ was achieved with again just a 2nd-order 
polynomial to reproduce the maximum of comoving space 
density around $z\simeq 2$. 
Alternatively, the sample can be described equally well by
pure luminosity evolution (PLE), albeit requiring one
more parameter than PDE. Our best-fit PLE model features
a 3rd order polynomial for the function $M^\star (z)$.
The predicted parametric luminosity 
functions are compared with the binned nonparametric estimates 
in Fig.\ \ref{fig:parametric}. Both fits provide excellent 
descriptions of the data, with goodness-of-fit probabilities
of $p > 50$\,\% for all statistical tests (see 
Tab.\ \ref{tab:parameters} for best-fit models). The differences 
between PDE and PLE luminosity functions become significant only 
outside the luminosity range sampled by the COMBO-17 data.

In Fig.\ \ref{fig:spacedens} we show the cumulative luminosity
function as a function of redshift, for different limiting 
luminosities, again together with the corresponding nonparametric 
estimates. For simplicity, only the PDE result is plotted.
Since the fit is presently unconstrained below $z=1.2$, we
have marked the extrapolation into this region by a dotted line.
The fit is completely consistent with the individual binned
datapoints.

\section{Discussion}

\subsection{The maximum of AGN activity}

The dominant feature in Fig.\ \ref{fig:spacedens} is
the peak of comoving AGN space densities around $z\simeq 2$.
Although its existence was beyond doubt for a long time, 
rarely has it been possible to locate the maximum 
\emph{within a single survey}. UV excess-selected samples 
typically reach up to $z = 2$--2.5 and mainly trace 
the rise at low $z$; conversely, most dedicated 
high-redshift surveys start being effective only 
beyond $z \simeq 3$ and just show the decreasing branch.
The COMBO-17 sample combines properties of both 
search techniques and covers enough redshift range 
that the peak of AGN activity is clearly bracketed.
The PDE and PLE fits both place the maximum at 
a redshift of $z_{\mathrm{max}}=2.1$.

An important issue for the astrophysical interpretation
of AGN evolution is a possible luminosity dependence of
the peak location. For example, in hierarchical structure 
formation one expects objects of higher mass objects to form later, 
which in a simple scenario of correlated masses and luminosities
should manifest in high-$L$ AGN to show a peak at lower redshifts. 
This is certainly not the case in our data; on the contrary, 
one might be tempted to speculate from Fig.\ \ref{fig:spacedens} 
that there could be a gradual shift of the maximum 
towards higher redshifts when the luminosity limit is increased. 

We have tried to model such a trend by including explicit 
luminosity-dependent density evolution parameters, 
e.g.\ in the form of a linear dependency between absolute magnitude 
and $z_{\mathrm{max}}$. While the data are statistically 
\emph{consistent} with a moderate shift, the best-fit model 
was always very close to simple PDE, despite the additional 
degrees of freedom. We conclude that
there is no indication for a dramatic difference between 
the space density peaks of AGN of different luminosities,
within the range covered by COMBO-17.

A similar but much stronger effect, a shift of maximum space density 
towards \emph{lower} redshift for very low luminosity AGN, has recently 
been reported for deep X-ray selected samples \citep{Cow03,Has03}. 
While a detailed comparison between the properties of
X-ray and optically selected AGN samples is
beyond the scope of this paper, we just note 
two aspects which need to be taken into account: Firstly,
the new X-ray selected samples probe even much deeper 
than COMBO-17 into 
the population of AGN with very low luminosities.
Assuming an Elvis et al. (1994) AGN spectrum, our
low-luminosity limit of around $M_{145} = -23$ corresponds to 
a 2--8~keV luminosity of roughly $L_x \simeq 43.6$; this is
just the point where the difference becomes visible,
according to Cowie et~al.\ and Hasinger et~al.

But secondly, the X-ray and optical luminosity functions of
high-redshift AGN also show significant dissimilarities. 
In particular, the X-ray LF of type~1 (broad-line) AGN 
near $\log L_x \simeq 44$ is very nearly constant
(cf.\ Cowie et al.; Hasinger et al.), while
the optical LF around the corresponding $M_{145} \simeq -24.6$
is still significantly rising (cf.\ Fig.\ \ref{fig:lf}).
This may be indicative of different spectral shapes between
low- and high-luminosity, or between low- and high-redshift AGN,
all leading to nontrivial $L_x$ to $M_{145}$ conversions.
Thus, while on one hand our $R<24$ sample may still be not quite
deep enough to see similar effects as found in the X-ray
domain, it is not even obvious that exactly the same effects 
should be expected for a yet deeper sample.

\begin{figure}
\includegraphics[clip,width=\hsize]{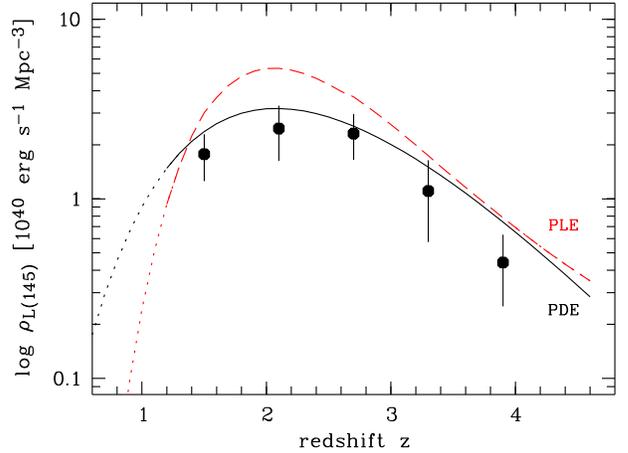}
\caption[]{Integrated UV luminosity density $\varrho_{L(145)}$ 
as a function of redshift, estimated from the cumulative LF
(data points) and for both PDE (solid) and PLE (dashed) models.
Below $z = 1.2$ the relations are extrapolated (dotted lines).
The discrepancy between PLE and PDE around the maximum 
is entirely due to the differences in the bright-end slope 
of the luminosity function.}
\label{fig:lumdens}
\end{figure}

\subsection{Evolution of the AGN luminosity density}

The contribution of AGN to the metagalactic radiation field
is probably relevant for several wavebands. Recent satellite missions
were very successful in resolving the extragalactic X-ray background
as mainly due to distant low-luminosity AGN \citep[e.g.][]{Miy00}.
Likewise, the diffuse ionising UV background is believed to be 
strongly influenced by AGN, but a quantitative synthesis still relies on
several assumptions and extrapolations. One of the principal
uncertainties has always been the shape of the low-luminosity end 
of the AGN luminosity function. In order to estimate the AGN 
luminosity density $\varrho_L(z)$, one has to evaluate the integral
\begin{equation}
  \varrho_L (z) = \int_L L\,\phi(L,z)\,dL \:.
\end{equation}
The integral diverges only for a luminosity function slope 
$\gamma < -2$, but the relative contributions of AGN at different
luminosity levels depend critically on the overall shape of the LF.

The COMBO-17 AGN sample goes deep enough that, for the first time, 
the quantity $L\,\phi$ can be safely integrated without depending on
heavy extrapolation into the unobserved range. We have computed
both binned nonparametric as well as parametric estimates of $\varrho_L(z)$,
which we show in Fig.\ \ref{fig:lumdens}. We present the monochromatic
luminosity density $\varrho_{L(145)}$, based on the value of 
$\lambda L_\lambda$ evaluated at $\lambda = 145$~nm
(which in turn is directly derived from $M_{145}$).
Assuming a typical QSO spectrum such as the one given by 
Elvis et al. (1994), the quantity 
$\varrho_{L(145)}$ can also be extrapolated into a 
frequency-integrated luminosity density. Since we are particularly
interested in the AGN contribution to the cosmic production of 
hydrogen-ionising photons, we provide the necessary conversion:
\begin{equation}
 L_{\lambda < 91.2\,\mathrm{nm}} \:=\: 
       1.45 \times \lambda L_{\lambda = 145\,\mathrm{nm}} \: .
\end{equation}
Notice that we have individual SED information on each of the 
COMBO-17 objects, which in principle could be used to derive
a more accurate estimate of $L_{\mathrm{ion}}$. This will be done
in a later paper dedicated to the synthesis of the 
UV background from the COMBO-17 AGN sample.

\begin{figure*}
\centering
\includegraphics[clip,angle=270,width=\hsize]{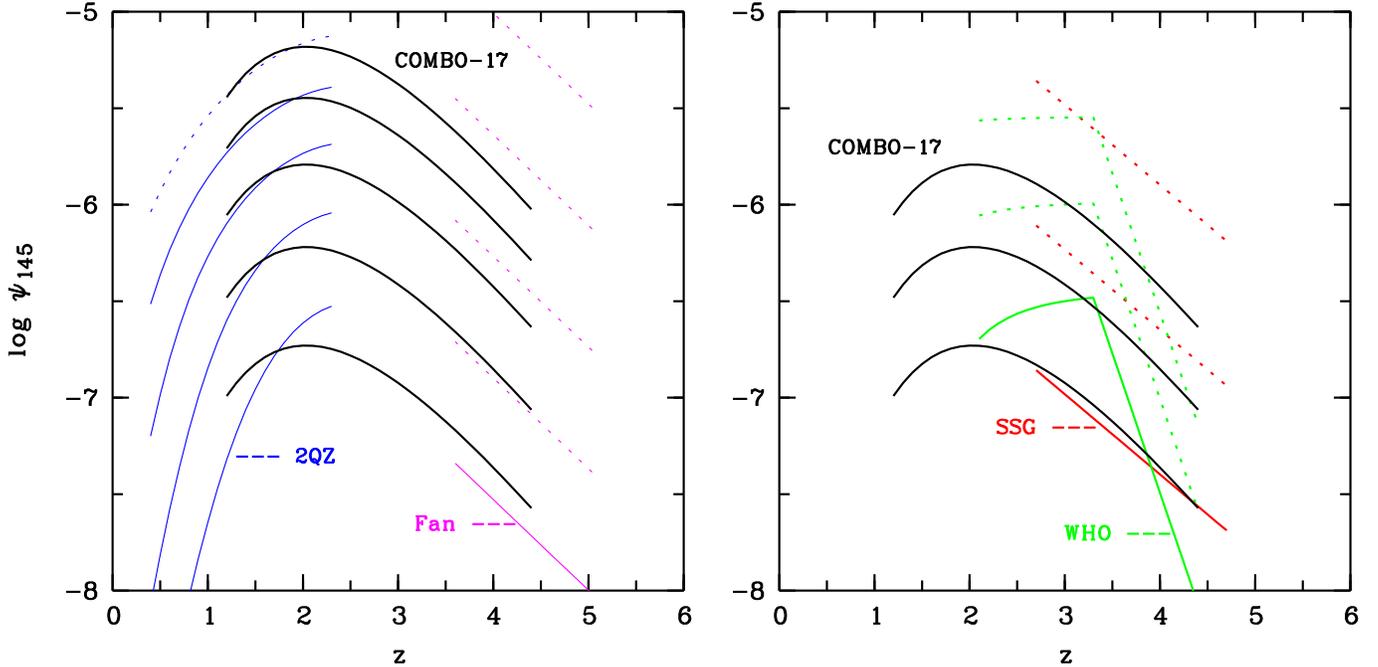}
\caption{Best-fitting parametric models of quasar space density in 
comparison: Solid lines are constrained by respective surveys and dashed 
lines are extrapolations to fainter luminosities.
{\it Left:} COMBO-17 with 2QZ and SDSS (Fan et al.), at luminosities of
$M_{145} < [-28 \ldots -24]$. The apparent disagreement of COMBO-17 
with 2QZ at brightest luminosities should not be taken seriously as the 
curves show the best-fitting model and not actual LF data. {\it Right:} 
COMBO-17 with WHO and SSG, at luminosities of $M_{145} < [-28,-27,-26]$.}
\label{allpsis}
\end{figure*}

For the nonparametric estimates we computed $\varrho_L$ by summing 
over the luminosity-weighted inverse volumes $L/V_i$ up to the 
survey limit; the resulting numbers should set lower limits to the 
full integral over all luminosities. 
The two adopted parametric forms of PDE and PLE can be directly
integrated to provide smooth functions $\varrho_L(z)$. It is worth noting 
that the integral effectively converges within the COMBO-17 survey limits;
the difference between setting the low-luminosity integration boundary 
at $M_{145} = -23$ (approximately the COMBO-17 limit)
and at $M_{145} = -10$ (or $L = 0$) is only 2\,\%. 
In other words, the constraints on the faint end of the AGN luminosity
function from COMBO-17 are already sufficient to estimate the total UV 
radiative output from AGN. This is also illustrated by the relatively
good agreement in Fig.\ \ref{fig:lumdens} 
between the binned estimates (which, as said above, 
are formally just lower limits to $\varrho_L$) and the PDE model.

On the other hand, the flat slope of the LF makes the contribution 
of brighter AGN to $\varrho_{L(145)}$ anything but negligible. 
In fact, the inclusion or exclusion of individual high-luminosity 
objects makes a noticeable difference for the binned estimate.
(This is much less so for the parametric estimates, because of the 
uniform weighting inherent in the maximum likelihood procedure.)
The importance of higher luminosity objects becomes particularly
apparent when comparing the PDE- and PLE-derived relations.
The substantial and significant differences seen in 
Fig.\ \ref{fig:lumdens} are \emph{not} due to faint-end extrapolation
effects, but entirely originate in the much flatter LF slope of the
PLE model at \emph{brighter} magnitudes, already documented 
in Fig.\ \ref{fig:parametric} above. We reiterate that these
discrepancies are due to the limited power of the COMBO-17 dataset
to discriminate between evolution models. 
They will be resolved as soon as constraints from other, 
brighter AGN surveys are combined with the COMBO-17 sample.

\subsection{Comparison with other surveys}

\subsubsection{Redshift $1.5<z<4.5$}

In terms of the luminosity range, our quasar sample pushes to fainter 
limits than any previous survey. Our targets mainly have observed
magnitudes of $R=20\ldots 24$, while previous surveys either observed down
to $R\la 20$ if they derived luminosity functions, or observed to $R\la 
22.5$ but produced object lists which did not constrain the luminosity 
functions very much further. Therefore, this work enters a new regime of
studying low-luminosity AGN at intermediate to high redshifts.

A first attempt to conduct an AGN survey with a strategy similar to
COMBO-17 was performed, albeit on a much smaller scale, in the course
of the CADIS survey (Wolf et al. 1999), where a sample of 12 QSOs at $z>2$
and $R<22$ was identified within an area of 250 arcmin$^2$.
Their resulting surface density is twice as high as for COMBO-17,
but because of the small sample size, the results of CADIS and
COMBO-17 are still formally compatible. The discrepancy underlines,
however, the importance of cosmic variance and hence the need to obtain
large samples.

Major previous work producing luminosity functions include the 2QZ at 
$z\la 2.3$ \citep{Boy00}, the work by WHO at $2.0<z<4.5$, and by SSG at 
$z>2.7$ as well as the results from the high-redshift SDSS sample at 
$z>3.6$ by Fan et al. (2001). As we have little overlap with all these 
previous surveys in luminosity, we cannot test directly to what extent 
our and their luminosity functions coincide.
However, at the bright end our LF is completely consistent with an 
extrapolation of the SDSS-based LF given by Fan et al., with similar slope 
and similar normalisation. SSG cover a slightly wider range of redshifts
down to $z=2.7$, but there still is not much overlap, so as before we need
to extrapolate. The SSG slope is steeper than that of Fan et al., so that 
the SSG prediction for our regime is even further above the COMBO-17 results.

\begin{figure}
\centering
\includegraphics[clip,angle=270,width=\hsize]{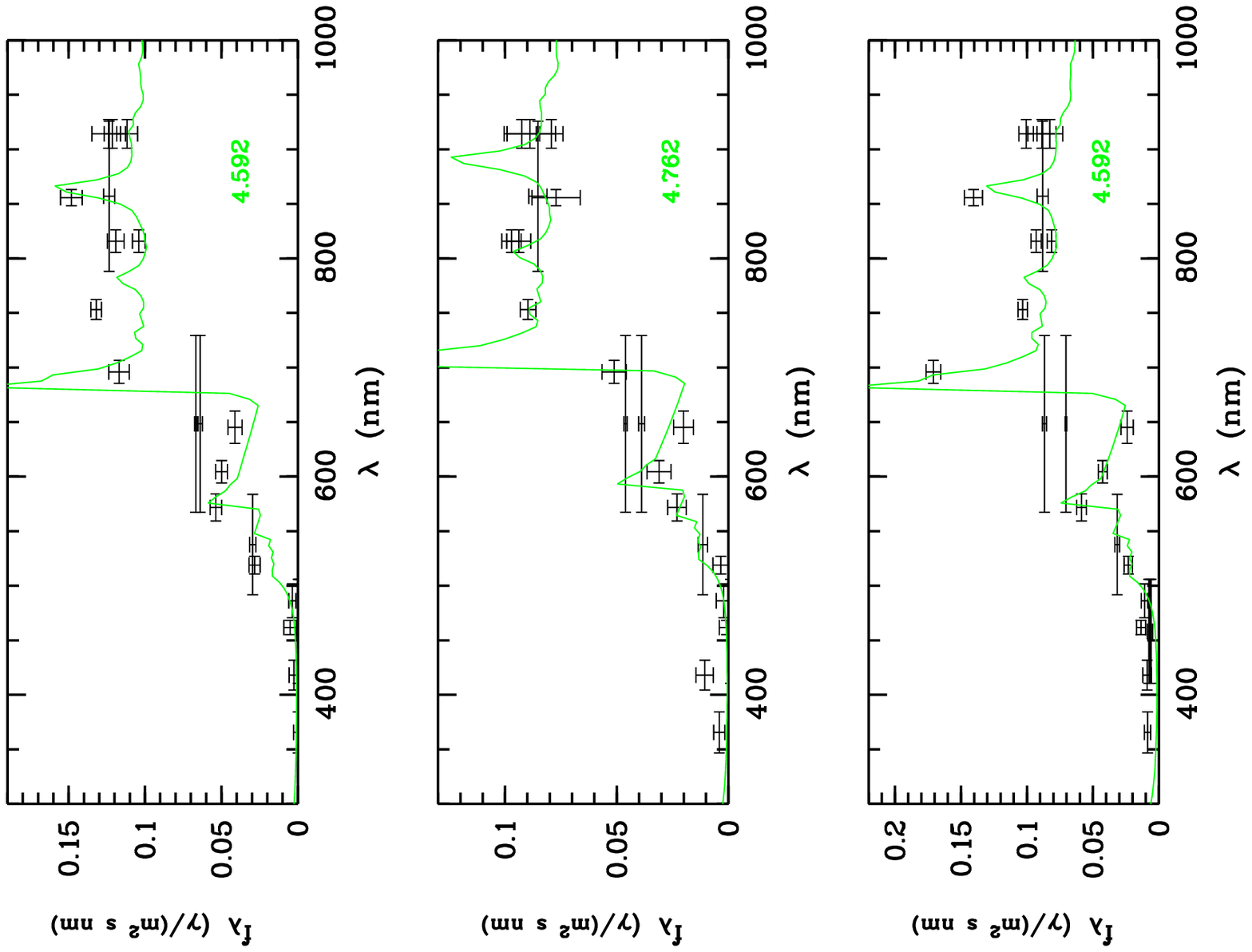}
\caption{Filter spectra of the three $z>4.5$ COMBO-17 quasars
in the present sample. See Fig.~\ref{examples} for interpretation
of the symbols.} 
\label{hizqsos}
\end{figure}

On the lower-redshift side, our results smoothly connect to the 2QZ results.
At intermediate redshift again, WHO used a broken power-law to characterize 
the QLF, where they found a steep end slope of $\alpha = -1.67$ and 
a faint end slope of $\beta = -0.45$, which is mostly constrained
by the probably non-optimal assumption of a broken power-law. 
The bright end of our LF varies around 0.4 to 0.7 in their units
and can be considered consistent with the WHO faint-end.

\subsubsection{Quasars at $z>4.5$}

The redshift range above 4.5 is subject to several studies from dedicated 
surveys to find faint high-redshift quasars. At highest luminosities the 
SDSS has delivered samples of six objects in 180~$\mbox{deg}^2$ at $i^*<20$
\citep{Fan01} and 29 objects in $\sim 700~\mbox{deg}^2$ at $i^*<20.5$ 
\citep{And01}. Medium-deep surveys across cover smaller areas, among which 
the BTC40 has so far reported two $z>4.5$ QSOs at $I<21.5$ across 
36~$\mbox{deg}^2$ and still have a list of fainter candidates at $I\la 22$ 
to follow up \citep{Mon02}. The Oxford-Dartmouth Thirty Degree Survey (ODT) 
also aims at finding $z>4.5$ quasars down to $I<22$ on 30~$\mbox{deg}^2$ 
(Dalton, priv. comm.). Monier et al. (2002) demonstrate the consistency of 
their currently available small numbers with the SDSS result and predict on
the basis of the SDSS-LF a surface density of 0.026~$/\mbox{deg}^2$ at 
$z>4.5\ldots 5.0$ and $I<19.9$, and much less at $z>5$.

COMBO-17 covers only 0.78~$\mbox{deg}^2$ currently, but reaches a bit deeper.
An extrapolation of the SDSS luminosity function suggests we should find 
2.0 quasars per $\mbox{deg}^2$ at $I<23$ and $z>4.5$. Our observation of some 
curvature in the fainter domain of the luminosity function reduces that 
prediction to $\approx 1.25/\mbox{deg}^2$, so one or two objects is the total 
number to be expected in the COMBO-17 dataset.

The present sample contains three objects at $z>4.5$, all in the S11 field 
(see Fig.~\ref{hizqsos} for filter spectra). They all have $I<22$, although 
our selection should be complete to $I<23$. Basically, this observation is
statistically consistent with an extrapolation of both the SDSS LF and our
own. As we can not draw strong conclusions about the cosmic abundance of 
these objects from COMBO-17, they have been excluded from the LF results
presented above.

\begin{table*}
\caption[]{Predicted AGN surface densities, based on (partly extrapolated) 
COMBO-17 results. Each field gives the cumulative AGN number per deg$^2$,
brighter than $R$ and with redshift greater than $z$, as derived from the 
two simple evolution models discussed in this paper.}
\label{tab:pred}
\begin{tabular}{c..@{\hspace{2em}}..@{\hspace{2em}}..@{\hspace{2em}}..}
\hline \hline
\noalign{\smallskip}
Magnitude & \multicolumn{2}{c}{$N(z>2)$\ \ } & \multicolumn{2}{c}{$N(z>3)$\ \ }
          & \multicolumn{2}{c}{$N(z>4)$ \ \ } & \multicolumn{2}{c}{$N(z>5)$ \ \ } \\
\noalign{\smallskip}
          & \multicolumn{1}{c}{PDE} & \multicolumn{1}{l}{\ PLE} & 
            \multicolumn{1}{c}{PDE} & \multicolumn{1}{l}{\ \ \ PLE} & 
            \multicolumn{1}{c}{PDE} & \multicolumn{1}{l}{\ \ PLE} & 
            \multicolumn{1}{c}{PDE} & \multicolumn{1}{l}{\ PLE} \\
\noalign{\smallskip}\hline\noalign{\smallskip}
$R < 20$  &  16.1 &  20.0  &   3.0 &  3.0   &   0.33 & 0.28  &  0.03 & 0.05 \\
$R < 21$  &  38.2 &  43.3  &   7.9 &  7.9   &   0.98 & 0.95  &  0.10 & 0.19 \\
$R < 22$  &  81.  &  85.   &  18.1 & 18.8   &   2.5  & 2.9   &  0.28 & 0.69 \\
$R < 23$  & 148.  & 152.   &  36.6 & 40.6   &   5.6  & 7.9   &  0.68 & 2.3  \\
$R < 24$  & 241.  & 252.   &  67.  & 81.    &  10.8  &19.7   &  1.5  & 6.7  \\
$R < 25$  & 322.  & 362.   & 101.  &147.    &  18.5  &44.8   &  2.7  &17.5  \\
\noalign{\smallskip}\hline
\end{tabular}
\end{table*}

\subsection{Predictions for future deep surveys}

Given a parametrised model for the evolving AGN luminosity function,
it is straightforward to predict surface densities (numbers
of AGN per unit solid angle) for other survey specifications.
In Table~\ref{tab:pred} we provide a set of such numbers,
computed for grid of $R$ band magnitudes and redshifts.
We deliberately stretch these predictions to the limit 
of credibility, e.g.\ in the lowest ($R < 25$) row or in the
rightmost ($z > 5$) column, in order to enable a direct comparison 
with possible future dedicated ultra-deep or very high redshift
surveys. It should be understood that some of these numbers involve
a considerable amount of extrapolation outside the areas
covered by our data. For the same reason we list at each grid point 
the predictions of both PDE and PLE models, hoping that the 
difference between these approximately bracket the level of 
uncertainty, especially in the extrapolation regions.

At intermediate redshifts and flux levels, well sampled
by COMBO-17 sources, the agreement of PDE and PLE is excellent,
simply reflecting the fact that both are valid descriptions
of the observed data. On the other hand, the two evolution modes
predict substantially different numbers of faint high-redshift 
AGN, the PLE prediction exceeding the PDE prediction by a factor
of several. This is only partly due to the uncertainties imposed
by the limited dataset. Figure \ref{fig:parametric} shows why this
effect is actually expected from the properties of the two 
chosen evolution modes: A luminosity function which fits the data 
well at $z\simeq 2$, which then is displaced either vertically or 
horizontally to account for the substantial negative evolution
beyond $z>3$, will result in dramatically different 
space densities of low-luminosity AGN.

\section{Conclusions}

We have presented work on the evolution of the quasar luminosity function
which has novel aspects by addressing two of the main unresolved problems 
in quasar research: The location of the peak of quasar activity, 
and the evolution of low-luminosity objects, 
the bulk of the quasar population: 
\begin{itemize}
\item Using the medium-band approach of COMBO-17 it was possible to break
 the colour degeneracies between stars and quasars around $z\sim 3$ which
 had prevented earlier optical surveys from mapping out the shape of the
 turnover in quasar activity from a single, homogeneous survey. As a 
 result, we have measured a broad maximum around $z\approx 2$.

\item By reaching faint magnitudes of $R<24$, we were probing the quasar
 LF and measured its slope down to rather low luminosity. The constraints
 on the faint end of the LF are fully sufficient to estimate
 the total UV radiative output. Although the two investigated evolution
 modes predict drastically different numbers of very faint ($R\ga 25$),
 very high redshift ($z\ga 5$) AGN, these have not much effect on the
 luminosity density integral. The main contribution to the UV background
 comes indeed from intermediate luminosity objects within $M_{145} \approx
 [-25,-28]$.
\end{itemize}

However, a number of questions and details remain still open.
Among these are:
\begin{enumerate}
\item The analysis of COMBO-17 catalogues themselves has so far omitted
 redshifts $z<1.2$, because low-luminosity Seyfert galaxies, also those of
 type 1, may remain undetected due to their complex SEDs consisting of a mix of
 contributions from the host galaxy and the active nucleus. It will be a
 technical issue for the COMBO-17 classification to improve on that in the
 near future.

\item At $z>4.5$ COMBO-17 reaches to sufficient depth, but does not cover
 sufficient area to make a significant contribution. Wider area surveys such
 as the BTC40, the ODT or a deep sub-area of SDSS are needed to identify a
 larger sample of faint high-redshift quasars.

\item Our sample alone fits well to both of the simplest evolutionary models
 for the quasar LF, the PDE and PLE models. Thus, further improvements in 
 determining the overall shape of the LF and understanding its evolution
 on a broader basis will require the combination of the COMBO-17 sample with
 other samples at higher luminosity and lower redshift (to be addressed by 
 a forthcoming paper).

\item While our survey is conducted in the rest-frame UV domain, one could
 wish to estimate the \emph{bolometric} luminosity density by applying a 
 bolometric correction factor. Using the Elvis et al. (1994) template SED,
 the relation would be $L_{\mathrm{bol}} = 4.5\times L_{145}$. However, we
 have little reason to assume \emph{a priori} that the quoted bolometric
 correction is universally applicable to all AGN including our faint sample,
 so at this stage we refrained from following this up further. For similar
 reasons we do not want to derive accretion rates here.

\item Probably, our sample breaks new ground in terms of depth and redshift
 coverage, however, it is clear that the results apply only to the evolution
 of broad-line type-1 AGN.

\item Deep X-ray surveys with the Chandra and XMM/Newton observatories
 have started to charter into new territory of AGN research. Comparing the
 first results of X-ray and deep optical samples such as COMBO-17, we
 see similarities but also a number of differences. 
 Exploring the interplay between X-ray and optical domain for AGN will be a
 highly rewarding exercise, which will lead to a more complete understanding
 of the cosmic evolution of accretion-powered galactic nuclei.
\end{enumerate}

In summary, understanding the accretion history of black holes in the nuclei 
of galaxies will certainly require some more quite fundamental work. However, 
establishing the evolution of faint quasars in the turnover epoch is one step 
forward.

\begin{acknowledgements}
CW was supported by the PPARC rolling grant in Observational Cosmology at 
University of Oxford and by the DFG--SFB 439. We thank Prof. Hasinger for 
providing spectroscopic redshifts of Chandra X-ray sources to facilitate the 
cross check with multi-colour redshifts and foster our trust in their quality. 
We thank Lance Miller, Steve Warren and an anonymous referee for helpful 
comments improving the manuscript.
\end{acknowledgements}

\bibliographystyle{aa}

\begin{thebibliography}{}
\bibitem[Andersen et al.\ 2001]{And01}
 Andersen, S. F., Fan, X., Richards, G. T., et al., 2001, AJ, 122, 503
\bibitem[Baade et al.\ 1998]{WFI1}
 Baade, D., Meisenheimer, K., Iwert, O., et al., 1998, The Messenger, 93, 13-15
\bibitem[Baum 1962]{Baum62}
 Baum, W. A., 1962, in {\it Problems of Extragalactic Research},
 ed. McVittie, G. C., Macmilliam, IAU Symposium 15, 390 
\bibitem[Bertin \& Arnouts 1996]{BA96}
 Bertin, E., Arnouts, S., 1996, A\&AS, 117, 393
\bibitem[Boyle et al.\ 2000]{Boy00}
 Boyle, B. J., Shanks, T., Croom, S. M., et al., 2000, MNRAS, 317, 1014
\bibitem[Boyle et al.\ 1988]{Boy88}
 Boyle, B. J., Shanks, T., Peterson, B. A., 1988, MNRAS, 235, 935
\bibitem[Boyle \& Terlevich 1998]{BT98}
 Boyle, B. J., Terlevich, R. J., 1998, MNRAS, 293, L49
\bibitem[Butchins 1983]{But83}
 Butchins, S. A., 1983, MNRAS, 203, 1239
\bibitem[Cowie et al.\ 2003]{Cow03}
 Cowie, L.L., Barger, A.J., Bautz, M.W., Brandt, W.N., Garmire, G.P.,
 2003, ApJ, 584, L57
\bibitem[Croom et al.\ 2001]{Cro01}
 Croom, S. M., Smith, R. J., Boyle, B. J., et al., 2001, MNRAS 322, L29
\bibitem[deBruijne et al.\ 2002]{dBr02}
 deBruijne, J. H. J., Reynolds, A. P., Perryman, M. A. C., et al., 2002,
 A\&A, 381, L57
\bibitem[Elvis et al.\ 1994]{Elv94}
 Elvis, M., Wilkes, B.J., McDowell, J.C., et al., 1994, ApJS, 95, 1
\bibitem[Fan et al.\ 2001]{Fan01}
 Fan, X., Strauss, M. A., Schneider, D. P., et al., 2001, AJ, 121, 54
\bibitem[Francis et al.\ 1991]{Fra91} Francis, P. J., Hewett, P. C.,
 Foltz, C. B., et al., 1991, ApJ, 373, 465
\bibitem[Haehnelt \& Rees 1993]{HR93}
 Haehnelt, M. G., Rees, M. J., 1993, MNRAS, 263, 168 
\bibitem[Haiman \& Loeb 1998]{Hai98}
 Haiman, Z., Loeb. A., 1998, ApJ, 503, 505
\bibitem[Hasinger et al.\ 2003]{Has03}
 Hasinger, G., and the CDFS team, 2003, in: {\it The Emergence of Cosmic 
 Structure}, S.S. Holt, C. Reynolds (eds.), in press, astro-ph/0302574
\bibitem[Irwin et al.\ 1991]{Irw91}
 Irwin, M., McMahon, R. G., Hazard, C., 1991, in {\it The Space 
 Distribution of Quasars}, Crampton, D. (ed.), ASP Conf. Ser. 21, p. 117
\bibitem[Jarvis \& Rawlings 2000]{JR00}
 Jarvis, M., J., Rawlings, S., 2000, MNRAS, 319, 121
\bibitem[Kauffmann \& Haehnelt 2000]{KH00}
 Kauffmann, G., Haehnelt, M. G., 2000, MNRAS, 311, 576
\bibitem[Kennefick et al.\ 1995]{Ken95}
 Kennefick, J. D., Djorgovski, S. G., deCarvalho, R. R., 1995, AJ 110, 2553
\bibitem[K\"ohler et al.\ 1997]{Koe97}
 K\"ohler, T., Groote, D., Reimers, D., Wisotzki, L., 1997, A\&A, 325, 502
\bibitem[Lilliefors et al.\ 1968]{Lil68}
 Lilliefors R., 1968, J.\ Am.\ Stat.\ Assoc.\ 62, 399
\bibitem[Marshall et al.\ 1983]{Mar83}
 Marshall, H. L., Avni, Y., Tananbaum, H., Zamorani, G., 1983, ApJ 269, 35
\bibitem[Meiksin \& Madau 1993]{MM93}
 Meiksin, A., Madau, P., 1993, ApJ, 412, 34
\bibitem[Miyaji et al.\ 2000]{Miy00}
 Miyaji, T., Hasinger, G., Schmidt, M., 2000, A\&A, 353, 25
\bibitem[Monier et al.\ 2002]{Mon02}
 Monier, E., Kennefick, J. D., Hall, P. B., et al., 2002, AJ, 124, 2971
\bibitem[Press et al.\ 1992]{Pre92}
 Press, W. H., Teukolsky, S. A., Vetterling, W.~T., Flannery, B.~P., 
 1992, Numerical Recipes in C, Cambridge University Press, 2nd edition
\bibitem[Richards et al.\ 2001]{Ric01}
 Richards, G. T., Weinstein, M. A., Schneider, D. P., et al., 2001,
 AJ, 122, 1151
\bibitem[R\"oser \& Meisenheimer 1991]{RM91} R\"oser, H.-J.,
 Meisenheimer, K., 1991, A\&A, 252, 458
\bibitem[Schlegel et al.\ 1998]{SFD98}
 Schlegel, D. J., Finkbeiner, D. P., Davies, M., 1998, ApJ, 500, 525
\bibitem[Schmidt 1968]{Sch68}
 Schmidt, M., 1968, ApJ, 151, 393
\bibitem[Schmidt et al.\ 1995]{SSG95}
 Schmidt, M., Schneider, D. P., Gunn, J. E., 1995, AJ, 110, 68
\bibitem[vanden Berk et al.\ 2001]{vdB01}
 vanden Berk, D. E., Richards, G. T., Bower, A., et al., 2001, AJ, 122, 549
\bibitem[Warren et al.\ 1994]{WHO94}
 Warren, S. J., Hewett, P. C., Osmer, P. S., 1994, ApJ, 421, 412
\bibitem[Wisotzki 1998]{Wis98}
 Wisotzki, L., 1998, AN, 319, 257
\bibitem[Wisotzki 2000]{Wis00}
 Wisotzki, L., 2000, A\&A, 353, 853
\bibitem[Wisotzki et al.\ 2000]{WC00}
 Wisotzki, L., Christlieb, N., Bade, N., et al., 2000, A\&A, 358, 77
\bibitem[Wolf et al.\ 1999]{Wol99}
 Wolf, C., Meisenheimer, K., R\"oser, H.-J., et al., 1999, A\&A, 343, 399 
\bibitem[Wolf et al.\ 2001a]{WMR01}
 Wolf, C., Meisenheimer, K., R\"oser, H.-J., 2001, A\&A, 365, 660
\bibitem[Wolf et al.\ 2001b]{Wol01}
 Wolf, C., Dye, S., Kleinheinrich, M., Meisenheimer, K., Rix, H.-W., Wisotzki,
 L., 2001, A\&A, 377, 442
\bibitem[Wolf et al.\ 2003]{Wol03}
 Wolf, C., Meisenheimer, K., Rix, H.-W., Borch, A., Dye, S., Kleinheinrich,
 M., 2003, A\&A, 401, 73
\end{thebibliography}

\end{document}